\documentclass{ajour}
\usepackage{latexsym}
\usepackage{feynmp}\usepackage[nosort]{cite}
\usepackage{feynmp-sup}
\usepackage{graphicx}
\def\MP{\mp}
\def\cal{\mathcal}
\def\intp{\int\frac{\di^4 p}{(2\pi)^4}}
\newcommand{\di}{{\mathrm d}}

\newcommand{\ii}{{\mathrm i}}

\renewcommand{\and}{\quad{\mathrm{and}}\quad}

\renewcommand{\Re}{{\mathrm{Re}}}
\renewcommand{\oint}{\int_{\cal C}}
\renewcommand{\Im}{{\mathrm{Im}}}
\def\Tc{{\cal T}_{\cal C}}

\def\scr#1{\mbox{\scriptsize #1}}

\def\vec#1{\mbox{\boldmath $#1$}}
\newcommand{\dpi}[1]{\frac{\di^4 #1}{(2\pi)^4}}                
\newcommand{\Pbr}[1]{\left\{#1\right\}}                    
\newlength{\charwidth}
\def\medhat#1{\settowidth{\charwidth}{$#1\,$}{\makebox[\charwidth]{$\,
 {\widehat{\makebox[2mm]{$#1\,$}}}$}}\vphantom{#1}}
\newcommand{\lap}%
{\raisebox{-0.5ex}{$\stackrel{\scriptstyle <}{\scriptstyle \sim}$}}
\newcommand{\gap}%
{\raisebox{-0.5ex}{$\stackrel{\scriptstyle >}{\scriptstyle \sim}$}}

\def\Gr{G}\def\Se{\Sigma}

\def\Sa{\Se}

\def\Lg{{\cal L}}
\def\Lgh{\makebox[3.5mm]{${\widehat{\makebox[2mm]{$\Lg$}}}$}\vphantom{L}}
\def\Lint{\Lgh^{\mbox{\scriptsize int}}}

\def\Gdashed{
\parbox{18\unitlength}{
\begin{fmfgraph*}(18,10)
\fmfleft{l}
\fmfright{r}
\fmf{dbl_plain,label=$i\hspace*{14\unitlength}j$,l.s=left}{r,l}
\end{fmfgraph*}
}}
\def\Pdashed#1#2{
\parbox{18\unitlength}{
\centerline{\begin{fmfgraph*}(18,10)
\fmfleft{l}
\fmfright{r}
\fmf{dashes_arrow,label=$#1\hspace*{14\unitlength}#2$,l.s=left}{r,l}
\end{fmfgraph*}}
}}
\def\FBox#1#2#3{ 
\parbox{19\unitlength}{
\begin{fmfgraph*}(19,26)
\fmfpen{thin}
\fmfleft{lb,lt}
\fmfright{rb,rt}
\fmf{phantom,tension=2}{rb,prb}
\fmf{phantom,tension=2}{lb,plb}
\fmf{phantom,tension=2}{rt,prt}
\fmf{phantom,tension=2}{lt,plt}
\fmfpoly{filled=30,label=#3}{prb,prt,plt,plb}
\fmfdot{plb,prb}
\fmflabel{#1}{plb}
\fmflabel{#2}{prb}
\end{fmfgraph*}
}}
\def\FBoxL{
\parbox{22\unitlength}{$ $\hfill
\begin{fmfgraph*}(19,26)
\fmfpen{thin}
\fmfleft{lb,lt}
\fmfright{rb,rt}
\fmf{phantom,tension=2}{rb,prb}
\fmf{phantom,tension=2}{lb,plb}
\fmf{phantom,tension=2}{rt,prt}
\fmf{phantom,tension=2}{lt,plt}
\fmfpoly{filled=30,label=$M'$}{prb,prt,plt,plb}
\fmffreeze
\fmf{dashes_arrow,left=0.8}{plb,plt}
\fmf{dbl_plain,left=0.8}{plt,prt}
\fmfdot{plb,prb}
\fmflabel{$1$}{plb}
\fmflabel{$2$}{prb}
\fmflabel{$3$}{plt}
\fmflabel{$4$}{prt}
\end{fmfgraph*}
}}
\def\FBoxR{
\parbox{22\unitlength}{
\begin{fmfgraph*}(19,26)
\fmfpen{thin}
\fmfleft{lb,lt}
\fmfright{rb,rt}
\fmf{phantom,tension=2}{rb,prb}
\fmf{phantom,tension=2}{lb,plb}
\fmf{phantom,tension=2}{rt,prt}
\fmf{phantom,tension=2}{lt,plt}
\fmfpoly{filled=30,label=$M'$}{prb,prt,plt,plb}
\fmffreeze
\fmf{dashes_arrow,right=0.8}{prb,prt}
\fmf{dbl_plain,left=0.8}{plt,prt}
\fmfdot{plb,prb}
\fmflabel{$1$}{plb}
\fmflabel{$2$}{prb}
\fmflabel{$3$}{plt}
\fmflabel{$4$}{prt}
\end{fmfgraph*}\hfill$ $
}}
\def\Ploop{\parbox{11\unitlength}{
\begin{fmfgraph*}(10,10)
\fmfpen{thin}
\fmftop{t}
\fmfleft{l}
\fmfbottom{b}
\fmf{dashes_arrow,left=1.0,tension=1}{t,b}
\fmf{dashes,left=1.0,tension=1}{b,t}
\fmfdot{l}
\end{fmfgraph*}
}}
\def\DRhomb#1#2{ \parbox{33\unitlength}
  {\centerline{\begin{fmfgraph**}(30,15,11) \fmfright{r} \fmfleft{l}
        \fmfforce{0.5w,0.5h}{m} \fmfpoly{filled=30,label=#1}{m,t1,l,b1}
        \fmfpoly{filled=30,label=#2}{m,b2,r,t2} \fmfdot{r,m,l}
\end{fmfgraph**}}}
}
\def\DPoisson{\parbox{33\unitlength}
{\centerline{\begin{fmfgraph**}(30,15,11)
\fmfright{r}
\fmfleft{l}
\fmfforce{0.5w,0.5h}{m}
\fmfpoly{filled=30,label=$A$}{m,t1,l,b1}
\fmfpoly{filled=30,label=$\partial_X B$}{m,b2,r,t2}
\fmf{dashes_arrow,left=1.3}{l,m}
\fmfdot{r,m,l}
\end{fmfgraph**}}}
+
\parbox{33\unitlength}
{\centerline{\begin{fmfgraph**}(30,15,11)
\fmfright{r}
\fmfleft{l}
\fmfforce{0.5w,0.5h}{m}
\fmfforce{0.25w,0.5h}{ll}
\fmfforce{0.75w,0.5h}{rr}
\fmfpoly{filled=30,l.d=0,label=$\partial_X A$}{m,t1,l,b1}
\fmfpoly{filled=30,label=$B$}{m,b2,r,t2}
\fmf{dashes_arrow,right=1.3}{r,m}
\fmfdot{r,m,l}
\end{fmfgraph**}}}
}
\def\Parachute{
\parbox{18\unitlength}{\centerline{
\begin{fmfgraph*}(15,25)\fmfkeep{parachute}
\fmfpen{thin}
\fmfleft{l}
\fmftop{t}
\fmfforce{(0.5w,0.2h)}{b}
\fmfright{r}
\fmf{phantom}{l,t2,t3,r}
\fmf{phantom,tension=0.2}{t2,t,t3}
\fmf{phantom,tension=0.2}{l,t2}
\fmf{phantom,tension=0.2}{t3,r}
\fmffreeze
\fmf{phantom,left=1,label=${\cal C}^{Dr}$}{t2,t3}
\fmf{plain,left=1}{l,t2,t3,r}
\fmf{plain,left=1.5}{l,r}
\fmf{fermion,label={$^{\bar{1}}$},l.side=left,l.d=0.9mm}{b,l}
\fmf{fermion,label={$^{\bar{2}}$},l.side=left,l.d=0.9mm}{b,t2}
\fmf{fermion,label={$^1$},l.side=left,l.d=0.9mm}{t3,b}
\fmf{fermion,label={$^2$},l.side=left,l.d=0.9mm}{r,b}
\fmfdot{b}
\fmflabel{$r^-$}{b}
\end{fmfgraph*}}
}}
\def\Pnu{
\parbox{11\unitlength}{
\centerline{
\begin{fmfgraph*}(8,8)
\fmfpen{thin}
\fmfleft{l}
\fmfright{r}
\fmf{dbl_wiggly,label=$\nu$,label.side=right}{r,l}
\fmfv{d.sh=hexagram,d.size=4thick}{l}
\end{fmfgraph*}}
}}
\def\PnuLr{
\parbox{11\unitlength}{
\centerline{
\begin{fmfgraph*}(8,16)
\fmfpen{thin}
\fmfleft{bl}
\fmfright{b}
\fmf{dbl_wiggly,label=$\nu$,label.side=left}{b,bl}
\fmfv{d.sh=hexagram,d.size=4thick}{bl}
\fmf{plain}{bl,b}
\fmf{dashes_arrow,left=2,label=$\mu$,label.side=right}{bl,b}
\end{fmfgraph*}}
}}
\def\PnuLl{
\parbox{11\unitlength}{
\centerline{
\begin{fmfgraph*}(8,16)
\fmfpen{thin}
\fmfleft{bl}
\fmfright{b}
\fmf{dbl_wiggly,label=$\nu$,label.side=left}{b,bl}
\fmfv{d.sh=hexagram,d.size=4thick}{bl}
\fmf{dashes_arrow,right=2,label=$\mu$,label.side=left}{b,bl}
\end{fmfgraph*}}
}}
\def\ParachL{
\parbox{24\unitlength}{
\centerline{
\begin{fmfgraph*}(18,30)
\fmfpen{thin}
\fmfleft{l}
\fmftop{t}
\fmfforce{(0.5w,0.2h)}{b}
\fmfright{r}
\fmf{phantom}{l,t2,t3,r}
\fmf{phantom,tension=0.2}{t2,t,t3}
\fmf{phantom,tension=0.2}{l,t2}
\fmf{phantom,tension=0.2}{t3,r}
\fmffreeze
\fmf{dbl_wiggly,label=$\nu$,label.side=left}{b,bl}
\fmfv{d.sh=hexagram,d.size=4thick}{bl}
\fmf{plain}{bl,l}
\fmf{phantom,left=0.6,label=${\cal C}^{Dr}$}{t2,t3}
\fmf{plain,left=1}{l,t2,t3,r}
\fmf{plain,left=1.5}{l,r}
\fmffreeze
\fmf{plain}{b,r}
\fmf{plain}{b,t2}
\fmf{plain}{t3,b}
\fmfdot{b}
\fmflabel{$\;\; r^-$}{b}
\end{fmfgraph*}
}}
}
\def\ParachR{
\parbox{24\unitlength}{
\centerline{
\begin{fmfgraph*}(18,30)
\fmfpen{thin}
\fmfleft{l}
\fmftop{t}
\fmfforce{(0.5w,0.2h)}{b}
\fmfright{r}
\fmf{phantom}{l,t2,t3,r}
\fmf{phantom,tension=0.2}{t2,t,t3}
\fmf{phantom,tension=0.2}{l,t2}
\fmf{phantom,tension=0.2}{t3,r}
\fmffreeze
\fmf{dbl_wiggly,label=$\nu$,label.side=right}{b,br}
\fmfv{d.sh=hexagram,d.size=4thick}{br}
\fmf{plain}{b,l}
\fmf{phantom,left=0.6,label=${\cal C}^{Dr}$}{t2,t3}
\fmf{plain,left=1}{l,t2,t3,r}
\fmf{plain,left=1.5}{l,r}
\fmf{plain}{br,r}
\fmffreeze
\fmf{plain}{b,t2}
\fmf{plain}{t3,b}
\fmfdot{b}
\fmflabel{$\;\; r^-$}{b}
\end{fmfgraph*}
}}
}
\def\ParachAll{
\parbox{21\unitlength}{
\centerline{
\begin{fmfgraph*}(18,30)
\fmfpen{thin}
\fmfleft{l}
\fmftop{t}
\fmfforce{(0.5w,0.2h)}{b}
\fmfright{r}
\fmf{phantom}{l,t2,t3,r}
\fmf{phantom,tension=0.2}{t2,t,t3}
\fmf{phantom,tension=0.2}{l,t2}
\fmf{phantom,tension=0.2}{t3,r}
\fmffreeze
\fmf{dbl_wiggly}{b,bl}
\fmfv{d.sh=hexagram,d.size=4thick}{bl}
\fmf{plain}{bl,l}
\fmf{phantom,left=0.6,label=${\cal C}$}{t2,t3}
\fmf{plain,left=1}{l,t2,t3,r}
\fmf{plain,left=1.5}{l,r}
\fmffreeze
\fmf{plain}{b,r}
\fmf{plain}{b,t2}
\fmf{plain}{t3,b}
\fmfdot{b}
\end{fmfgraph*}
}}
}
\def\Paracha{
\parbox{23\unitlength}{
\begin{fmfgraph*}(18,30)
\fmfpen{thin}
\fmfleft{l}
\fmftop{t}
\fmfforce{(0.5w,0.2h)}{b}
\fmfright{r}
\fmf{phantom}{l,t2,t3,r}
\fmf{phantom,tension=0.2}{t2,t,t3}
\fmf{phantom,tension=0.2}{l,t2}
\fmf{phantom,tension=0.2}{t3,r}
\fmffreeze
\fmf{phantom,left=0.6,label=${{\cal C}'}$}{t2,t3}
\fmf{plain,left=1}{l,t2,t3,r}
\fmfforce{(0.2w,0.9h)}{ll}
\fmfforce{(0.8w,0.9h)}{rr}
\fmf{plain,left=.3}{l,ll,rr,r}
\fmffreeze
\fmf{dbl_wiggly}{b,bl}
\fmf{plain}{bl,l}
\fmfv{d.sh=hexagram,d.size=4thick}{bl}
\fmf{plain}{b,t2}
\fmf{plain}{t3,b}
\fmf{plain}{r,b}
\fmf{dbl_plain,left=1}{ll,rr}
\fmf{dashes_arrow,right=1}{b,rr}
\fmfdot{b}
\fmflabel{(a)}{b}
\end{fmfgraph*}
}}
\def\Parachb{
\parbox{23\unitlength}{\hspace*{4\unitlength}
\begin{fmfgraph*}(18,30)
\fmfpen{thin}
\fmfleft{l}
\fmftop{t}
\fmfforce{(0.5w,0.2h)}{b}
\fmfright{r}
\fmf{phantom}{l,t2,t3,r}
\fmf{phantom,tension=0.2}{t2,t,t3}
\fmf{phantom,tension=0.2}{l,t2}
\fmf{phantom,tension=0.2}{t3,r}
\fmffreeze
\fmf{dbl_wiggly}{b,bl}
\fmf{plain}{bl,l}
\fmfv{d.sh=hexagram,d.size=4thick}{bl}
\fmf{phantom,left=0.6,label=${{\cal C}'}$}{t2,t3}
\fmf{plain,left=1}{l,t2,t3,r}
\fmfforce{(0.2w,0.9h)}{ll}
\fmfforce{(0.8w,0.9h)}{rr}
\fmf{plain,left=.3}{l,ll,rr,r}
\fmffreeze
\fmf{plain}{b,t2}
\fmf{plain}{t3,b}
\fmf{plain}{r,b}
\fmf{dbl_plain,left=1}{ll,rr}
\fmf{dashes_arrow,left=1}{b,ll}
\fmfdot{b}
\fmflabel{(b)}{b}
\end{fmfgraph*}
}
}
\def\Parachc{
\parbox{22\unitlength}{
\centerline{
\begin{fmfgraph*}(18,30)
\fmfpen{thin}
\fmfleft{l}
\fmftop{t}
\fmfforce{(0.5w,0.2h)}{b}
\fmfright{r}
\fmf{phantom}{l,t2,t3,r}
\fmf{phantom,tension=0.2}{t2,t,t3}
\fmf{phantom,tension=0.2}{l,t2}
\fmf{phantom,tension=0.2}{t3,r}
\fmffreeze
\fmf{dbl_wiggly}{b,bl}
\fmfv{d.sh=hexagram,d.size=4thick}{bl}
\fmf{dbl_plain}{bl,l}
\fmf{phantom,left=0.6,label=${\cal C}$}{t2,t3}
\fmf{plain,left=1}{l,t2,t3,r}
\fmf{plain,left=1.5}{l,r}
\fmffreeze
\fmf{plain,label=$ $}{b,r}
\fmf{dashes_arrow,left=1.8}{b,bl}
\fmf{plain}{b,t2}
\fmf{plain}{t3,b}
\fmfdot{b}
\fmflabel{(c)}{b}
\end{fmfgraph*}}
}}
\def\Parachd{
\parbox{22\unitlength}{
\centerline{
\begin{fmfgraph*}(18,30)
\fmfpen{thin}
\fmfleft{l}
\fmftop{t}
\fmfforce{(0.5w,0.2h)}{b}
\fmfright{r}
\fmf{phantom}{l,t2,t3,r}
\fmf{phantom,tension=0.2}{t2,t,t3}
\fmf{phantom,tension=0.2}{l,t2}
\fmf{phantom,tension=0.2}{t3,r}
\fmffreeze
\fmf{dbl_wiggly}{b,bl}
\fmfv{d.sh=hexagram,d.size=4thick}{bl}
\fmf{dbl_plain}{bl,l}
\fmf{phantom,left=0.6,label=${\cal C}$}{t2,t3}
\fmf{plain,left=1}{l,t2,t3,r}
\fmf{plain,left=1.5}{l,r}
\fmffreeze
\fmf{plain,label=$ $}{b,r}
\fmf{dashes_arrow,left=1}{b,l}
\fmf{plain}{b,t2}
\fmf{plain}{t3,b}
\fmfdot{b}
\fmflabel{(d)}{b}
\end{fmfgraph*}}
}}
\def\Parachf{
\parbox{20\unitlength}{
\centerline{
\begin{fmfgraph*}(18,30)
\fmfpen{thin}
\fmfleft{l}
\fmftop{t}
\fmfforce{(0.5w,0.2h)}{b}
\fmfright{r}
\fmf{phantom}{l,t2,t3,r}
\fmf{phantom,tension=0.2}{t2,t,t3}
\fmf{phantom,tension=0.2}{l,t2}
\fmf{phantom,tension=0.2}{t3,r}
\fmffreeze
\fmf{dbl_wiggly}{b,bl}
\fmfv{d.sh=hexagram,d.size=4thick}{bl}
\fmf{plain}{bl,l}
\fmf{phantom,left=0.6,label=${\cal C}$}{t2,t3}
\fmf{plain,left=1}{l,t2,t3,r}
\fmf{plain,left=1.5}{l,r}
\fmffreeze
\fmf{dbl_plain}{b,r}
\fmf{dashes_arrow,right=1}{b,r}
\fmf{plain}{b,t2}
\fmf{plain}{t3,b}
\fmfdot{b}
\fmflabel{(f)}{b}
\end{fmfgraph*}}
}}
\def\Parachalpha{
\parbox{27\unitlength}{
\centerline{
\begin{fmfgraph*}(18,30)
\fmfpen{thin}
\fmfleft{l}
\fmftop{t}
\fmfforce{(0.5w,0.2h)}{b}
\fmfright{r}
\fmf{phantom}{l,t2,t3,r}
\fmf{phantom,tension=0.2}{t2,t,t3}
\fmf{phantom,tension=0.2}{l,t2}
\fmf{phantom,tension=0.2}{t3,r}
\fmffreeze
\fmf{dbl_wiggly}{b,bl}
\fmfv{d.sh=hexagram,d.size=4thick}{bl}
\fmf{plain}{bl,l}
\fmf{phantom,left=0.6,label=$\partial_X{\cal C}$}{t2,t3}
\fmf{plain,left=1}{l,t2,t3,r}
\fmf{plain,left=1.5}{l,r}
\fmffreeze
\fmf{plain,label=$ $}{b,r}
\fmf{dashes_arrow,right=1}{b,r}
\fmf{plain}{b,t2}
\fmf{plain}{t3,b}
\fmfdot{b,r}
\fmflabel{\bf f}{r}
\fmflabel{($\alpha$)}{b}
\end{fmfgraph*}\hspace*{2mm}
}}}
\def\Parachbeta{
\parbox{25\unitlength}{
\centerline{
\begin{fmfgraph*}(18,30)
\fmfpen{thin}
\fmfleft{l}
\fmftop{t}
\fmfforce{(0.5w,0.2h)}{b}
\fmfright{r}
\fmf{phantom}{l,t2,t3,r}
\fmf{phantom,tension=0.2}{t2,t,t3}
\fmf{phantom,tension=0.2}{l,t2}
\fmf{phantom,tension=0.2}{t3,r}
\fmffreeze
\fmf{dbl_wiggly}{b,bl}
\fmfv{d.sh=hexagram,d.size=4thick}{bl}
\fmf{dbl_plain}{bl,l}
\fmf{phantom,left=0.6,label=${\cal C}$}{t2,t3}
\fmf{plain,left=1}{l,t2,t3,r}
\fmf{plain,left=1.5}{l,r}
\fmffreeze
\fmf{plain,label=$ $}{b,r}
\fmf{dashes_arrow,left=1.8}{b,bl}
\fmf{plain}{b,t2}
\fmf{plain}{t3,b}
\fmfdot{b}
\fmflabel{($\beta$)}{b}
\end{fmfgraph*}}
}}
\def\ParachEint{
\parbox{24\unitlength}{
\centerline{
\begin{fmfgraph*}(18,30)
\fmfpen{thin}
\fmfleft{l}
\fmftop{t}
\fmfforce{(0.5w,0.2h)}{b}
\fmfright{r}
\fmf{phantom}{l,t2,t3,r}
\fmf{phantom,tension=0.2}{t2,t,t3}
\fmf{phantom,tension=0.2}{l,t2}
\fmf{phantom,tension=0.2}{t3,r}
\fmffreeze
\fmf{plain}{b,l}
\fmf{phantom,left=0.6,label=${\cal C}^{Dr}$}{t2,t3}
\fmf{plain,left=1}{l,t2,t3,r}
\fmf{plain,left=1.5}{l,r}
\fmffreeze
\fmf{plain}{b,r}
\fmf{plain}{b,t2}
\fmf{plain}{t3,b}
\fmfdot{b}
\fmflabel{$\;\;\; r^-$}{b}
\end{fmfgraph*}
}}}
\def\PhiHartreeT{
\parbox{10\unitlength}{\centerline{
\begin{fmfgraph*}(10,20)
\fmftop{t}\fmfbottom{b}
\fmf{phantom}{t,tt,bb,b}\fmffreeze
\fmf{plain,right=1}{tt,t}
\fmf{fermion,right=1}{t,tt}
\fmf{plain,right=1}{b,bb}
\fmf{fermion,right=1}{bb,b}
\fmf{dashes}{bb,tt}
\end{fmfgraph*}}}}
\def\SigmaHartreeT{
\parbox{10\unitlength}{\centerline{
\begin{fmfgraph*}(10,15)
\fmfstraight
\fmftop{rt,t,lt}\fmfbottom{rbb,bb,lbb}
\fmf{phantom}{t,tt,bb}\fmffreeze
\fmf{plain,right=1}{tt,t}
\fmf{fermion,right=1}{t,tt}
\fmf{fermion}{bb,rrbb}\fmf{plain}{bb,llbb}
\fmf{phantom,tension=2}{rbb,rrbb}\fmf{phantom,tension=2}{lbb,llbb}
\fmf{dashes}{bb,tt}
\end{fmfgraph*}}}}
\def\PhiRingT#1{
\parbox{23\unitlength}{\centerline{
\begin{fmfgraph*}(20,20)
\fmfsurroundn{e}{#1}
\fmfcyclen{fermion,right=0.25}{e}{#1}
\begin{fmffor}{n}{1}{1}{#1}
\fmf{dashes}{e[n],i[n]}
\end{fmffor}
\fmfcyclen{fermion,right=0.25,tension=1.5}{i}{#1}
\end{fmfgraph*}}}}
\def\PhiHartree{
\parbox{15\unitlength}{\centerline{
\begin{fmfgraph*}(10,20)
\fmftop{t}\fmfbottom{b}\fmf{phantom}{t,m,b}
\fmf{fermion,right=1}{t,m}\fmf{fermion,left=1}{m,b}
\fmf{plain,right=1}{m,t}\fmf{plain,left=1}{b,m}
\end{fmfgraph*}}}}
\def\SigmaHartree{
\parbox{15\unitlength}{\centerline{
\begin{fmfgraph*}(10,20)
\fmftop{t}\fmfbottom{b}\fmf{phantom}{t,m,b}
\fmffreeze
\fmf{fermion,right=1}{t,m}
\fmf{plain,right=1}{m,t}
\fmfleft{l}\fmfright{r}
\fmf{phantom}{l,ll}\fmf{phantom}{rr,r}
\fmf{plain,tension=0.2}{rr,m,ll}
\fmfdot{m}
\end{fmfgraph*}}}}
\def\EintHartree{
\parbox{15\unitlength}{\centerline{
\begin{fmfgraph*}(10,20)
\fmftop{t}\fmfbottom{b}\fmf{phantom}{t,m,b}
\fmf{fermion,right=1}{t,m}\fmf{fermion,left=1}{m,b}
\fmf{plain,right=1}{m,t}\fmf{plain,left=1}{b,m}
\fmfdot{m}
\end{fmfgraph*}}}}
\def\PhiSandwich{
\parbox{25\unitlength}{\centerline{
\begin{fmfgraph*}(18,18)
\fmfleft{ll}\fmfright{rr}
\fmf{fermion,right=0.9}{rr,ll}\fmf{fermion,left=0.35}{rr,ll}
\fmf{fermion,right=0.9}{ll,rr}\fmf{fermion,left=0.35}{ll,rr}
\end{fmfgraph*}}}}
\def\EintSandwich{
\parbox{25\unitlength}{\centerline{
\begin{fmfgraph*}(18,18)
\fmfleft{ll}\fmfright{rr}
\fmf{fermion,right=0.9}{rr,ll}\fmf{fermion,left=0.35}{rr,ll}
\fmf{fermion,right=0.9}{ll,rr}\fmf{fermion,left=0.35}{ll,rr}
\fmfdot{ll}
\end{fmfgraph*}}}}
\def\SigmaSandwich{
\parbox{25\unitlength}{\centerline{
\begin{fmfgraph*}(18,18)
\fmfleft{l}\fmfright{r}
\fmfforce{(0.1w,0.5h)}{ll}\fmfforce{(0.9w,0.5h)}{rr}
\fmf{plain}{l,ll}\fmf{plain}{rr,r}
\fmf{fermion,right=0.7}{rr,ll}\fmf{fermion}{rr,ll}
\fmf{fermion,right=0.7}{ll,rr}
\fmfdot{ll}\fmfdot{rr}
\end{fmfgraph*}}}}
\def\PhiRing#1{
\parbox{23\unitlength}{\centerline{
\begin{fmfgraph*}(20,20)
\fmfsurroundn{e}{#1}
\fmfcyclen{fermion,right=0.25}{e}{#1}
\fmfrcyclen{fermion,right=0.25}{e}{#1}
\end{fmfgraph*}}}}
\def\EintRing#1{
\parbox{23\unitlength}{\centerline{
\begin{fmfgraph*}(20,20)
\fmfsurroundn{e}{#1}
\fmfcyclen{fermion,right=0.25}{e}{#1}
\fmfrcyclen{fermion,right=0.25}{e}{#1}
\fmfdot{e[#1/2+1]}
\end{fmfgraph*}}}}
\def\SigmaRing#1{
\parbox{30\unitlength}{\centerline{
\begin{fmfgraph*}(20,25)
\fmfsurroundn{e}{2*#1}
\fmfn{fermion,right=0.25}{e}{#1+1}
\begin{fmffor}{n}{1}{1}{#1}
\fmf{fermion,right=0.25}{e[n+1],e[n]}
\end{fmffor}
\fmffreeze
\fmfforce{(-0.2w,0.5h)}{l}
\fmfforce{(1.2w,0.5h)}{r}
\fmf{fermion}{e[1],e[#1+1]}
\fmf{plain}{r,e[1]}\fmf{plain}{e[#1+1],l}
\fmfdot{e[#1+1]}\fmfdot{e[1]}
\end{fmfgraph*}}}}
\def\SigmaRingT#1{
\parbox{28\unitlength}{\centerline{
\begin{fmfgraph*}(25,18)
\fmfstraight
\fmfleft{lb,l,lt}
\fmfright{rb,r,rt}
\fmfn{phantom}{b}{#1}
\fmf{phantom}{rb,b[1]}
\fmf{phantom}{b[#1],lb}
\fmfn{phantom}{m}{#1}
\fmf{phantom}{r,m[1]}
\fmf{phantom}{m[#1],l}
\fmffreeze
\fmf{fermion}{rb,lb}
\fmf{fermion}{m[1],m[#1]}
\begin{fmffor}{n}{1}{1}{#1}
\fmf{dashes}{m[n],b[n]}
\end{fmffor}
\fmf{fermion,left=1}{m[#1],m[1]}
\end{fmfgraph*}}}}
\def\contourxy{
\parbox{17\unitlength}{
\begin{picture}(17,2.)\thicklines
\put(0,1){
\contour
\put(14.5,-.5){\makebox(0,0){$\infty$}}
\put(11,-.5){\makebox(0,0){$t_x^+$}}
\put(8,1.8){\makebox(0,0){$t_y^-$}}
\put(11,0){\circle*{.2}}
\put(8,1){\circle*{.2}}
}
\end{picture}}}
\def\contour{\thicklines
\put(1,-.5){\makebox(0,0){$t_{0}$}}
\put(16.5,.5){\makebox(0,0){$t$}}
\put(0,.5){\vector(1,0){16}}
\put(1,1.){\line(1,0){13}}\put(14,.5){\oval(1,1)[br]}
\put(14,0){\vector(-1,0){13}}\put(14,.5){\oval(1,1)[tr]}}

\begin{document}
\authorrunninghead{J. Knoll et al.}
\titlerunninghead{Exact Conservation Laws}
\title{Exact Conservation
Laws  of the  Gradient Expanded Kadanoff--Baym Equations}

\author{J. Knoll$^{1}$, Yu. B. Ivanov$^{1,2}$ 
and D. N. Voskresensky$^{1,3}$} 


\affil{ $^{1}${\it Gesellschaft f\"ur Schwerionenforschung mbH, Planckstr. 1,
64291 Darmstadt, Germany} \\ $^{2}${\it Kurchatov Institute, Kurchatov sq. 1,
Moscow 123182, Russia} \\ $^{3}${\it Moscow Institute for Physics and
Engineering, Kashirskoe sh. 31, Moscow 115409, Russia} }
\email{J.Knoll@gsi.de, Y.Ivanov@gsi.de, D.Voskresensky@gsi.de} 

\abstract{ It is shown that the Kadanoff--Baym equations at consistent
first-order gradient approximation reveal exact rather than approximate
conservation laws related to global symmetries of the system. The conserved
currents and energy--momentum tensor coincide with corresponding Noether
quantities in the local approximation.  These exact conservations are valid,
provided a $\Phi$ derivable approximation is used to describe the system, and
possible memory effects in the collision term are also consistently evaluated
up to first-order gradients.}
\begin{article}
\begin{fmffile}{exact-cons-fig}
\fmfset{thin}{1.3pt}
\fmfset{arrow_len}{2.5mm}
\section{Introduction}
Non-equilibrium Green function techniques, developed by Schwinger,
Kadanoff, Baym and Keldysh \cite{Schw,Kad62,Keld64,LP}, provide the
appropriate concepts to study the space--time evolution of many-particle
quantum systems. This formalism finds now applications in various fields, such
as quantum chromo-dynamics \cite{Land,BI00}, nuclear physics, in particular
heavy ion collisions
\cite{Dan84,Dan90,Toh,Bot90,MSTV,Vos93,Hen,Knoll95,IKV,Bozek98,Knoll98,IKV99,Cass99,Leupold,Mosel,IKHV00,HK00},
astrophysics \cite{MSTV,VS87,Keil}, cosmology \cite{CalHu}, spin systems
\cite{Manson}, lasers \cite{Korenman}, physics of plasma \cite{Bez,Kraft},
physics of liquid $^{3}$He \cite{SerRai}, critical phenomena, quenched random
systems and disordered systems \cite{Chou}, normal metals and super-conductors
\cite{VS87,Rammer,Fauser}, semiconductors \cite{LipS}, tunneling and secondary
emission \cite{Noziers}, etc.

For actual calculations certain approximation steps are necessary.  In many
cases perturbative approaches are insufficient, like for systems with strong
couplings as treated in nuclear physics.  In such cases, one has to resum
certain sub-series of diagrams in order to obtain a reasonable approximation
scheme. In contrast to perturbation theory, for such resummations one
frequently encounters the fact that the scheme may no longer be conserving,
although for each diagram considered the conservation laws are implemented at
each vertex. Thus, the resulting equations of motion may no longer comply with
the conservation laws, e.g., of currents, energy and momentum. This is a
problem of particular importance for recent studies of particles with broad
damping width such as resonances \cite{IKV99}.  The problem of conservation
laws in such resummation schemes has first been considered in two pioneering
papers by Baym and Kadanoff \cite{KadB,Baym} discussing the response to an
external perturbation of quantum systems in thermodynamic equilibrium. Baym,
in particular, showed \cite{Baym} that any approximation, in order to be
conserving, must be based on a generating functional $\Phi$.  This functional
was first considered by Luttinger and Ward \cite{Luttinger} in the context of
the thermodynamic potential, cf. \cite{Abrikos}, and later reformulated in
terms of path integrals \cite{CJT}. For truncated self-consistent Dyson
resummations this functional method provides conserved Noether currents and
the conservation of total energy and momentum at the expectation value
level. In our previous paper \cite{IKV} we extended the concept to the
real-time Green function technique and relativistic systems, constructing
conserved 4-currents and local energy--momentum tensor for any chosen
approximation to the $\Phi$ functional.  While $\Phi$-derivable Dyson
resummations formulated in terms of the integro-differential Kadanoff-Baym
(KB) equations indeed provide {\em exact} conservation laws, these equations
are usually not directly solvable. Therefore many applications involve further
approximations to the KB equations: the {\em gradient} approximation and often
the {\em quasi-particle} approximation leading to differential equations of
mean field and transport type. Any improvement of the quasi-particle
approximation beyond the mean-field level, e.g., through inclusion of
energy-momentum-dependent self-energies, again leads to difficulties with
conservation laws. Various attempts to remedy this problem were undertaken,
see refs.  \cite{LipS,LSV,Bonitz,Bornath,SCFNW,Jeon,VBRS} and references
therein. An essential progress within the quasi-particle approximation was
achieved by the ans\"atze of refs. \cite{LipS,Bornath}.

Interested in the dynamics of particles with broad mass width like resonances
we like to discuss the question of conservation laws for transport problems at
a much more general level, than usually considered. We call this the {\em
quantum transport} level. It completely avoids the quasi-particle
approximation, and solely rests on the first-order gradient approximation of
the KB equations. This concept was first addressed by Kadanoff and Baym
\cite{Kad62} in the chapter ``Slowly varying disturbances'',
Eqs. (9-25). Motivated by applications for the description of heavy-ion
collisions further attempts were recently suggested
\cite{IKV99,Cass99,Leupold,Mosel,IKHV00,HK00} based on the so called
Botermans--Malfliet (BM) substitution \cite{Bot90}.  Within the BM choice of
the quantum kinetic equations a number of desired properties including an
H-theorem for a local entropy current related to these equations were derived
\cite{IKV99}, however no strict realization of the conservation laws for the
Noether currents were obtained\cite{IKV99,Leupold}. Due to the approximation
steps involved one may expect the quantum transport equations both, in the KB
and BM forms, to possess only approximate conservation laws though in line with
the level of approximation. Such approximate nature of conservation laws may
be well acceptable theoretically. Nevertheless, both from a principle
perspective and also from a practical point of view this situation is less
satisfactory. Quantum kinetic equations which possess exact conservation laws
related to the symmetries of the problem could serve as a natural extension of
the quasiparticle transport phenomenology to broad resonances, e.g. applicable
to high energy heavy ion collisions.

In this paper we give a proof that the quantum kinetic equations in the form
originally derived by Kadanoff and Baym in fact possess the generic feature of
exact conservation laws at the expectation value level. This holds provided
all self-consistent self-energies are generated from a $\Phi$-functional and
all possible memory effects due to internal vertices within the self-energy
diagrams are also consistently expanded to first-order gradients.

In sect. \ref{QKE} we review the derivation of the quantum kinetic equations
in a notation suitable for our later derivation of the conservation laws in
sect. \ref{CL}. Sect. \ref{Gadient-Apprx} deals with the general gradient
approximation and its representation in terms of diagrams. Finally, in sect.
\ref{Phi-Conserv} within the $\Phi$-derivable method we give a diagrammatic
proof of the exact conservation laws of the quantum kinetic equations for the
KB choice.
 
We restrict the presentation to physical systems described by complex quantum
fields of different constituents interacting via local couplings. The
kinematics can be either relativistic or non-relativistic. Extension to
real boson fields, as well as to relativistic fermions is straight forward
though tedious in the latter case. We also exclude theories with derivative
couplings. 
\unitlength6mm
\begin{figure}[b]
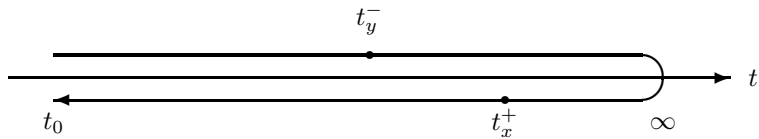

\begin{center}
\contourxy
\end{center}
\caption{\label{Fcontour}Closed real-time contour with two external points
$x,y$.} 
\end{figure}

\section{Kadanoff--Baym equations and complete gradient approximation}
\label{QKE}
We assume the reader to be familiar with the real-time formulation of
non-equilibrium field theory on the so called closed time contour,
Fig. \ref{Fcontour}. Since we will deal with general multi-point functions
we use the more convenient $\{-+\}$ contour-vertex notation of refs.
\cite{LP,IKV99}. Some rules are summarized in Appendix \ref{Contour}.  The set
of coupled KB equations on the time contour in $-+$ notation\footnote{The
numbers 1, 2 and 3 provide short hand notation for space--time coordinates
$x_1,x_2$ and $x_3$, respectively, including internal quantum numbers. With
superscript, like $1^-$ and $2^+$, assigned to them, they denote contour
coordinates with $-$ and $+$ specifying the placement on the time or anti-time
ordered branch. Decomposed to the two branches the contour functions are
denoted as $F^{kl}(1,2)=F(1^k,2^l)$ with $k,l\in \{-,+\}$. The match to the
notation used, e.g., in refs. \cite{Kad62,Dan84} is given by $F^{-+}=F^{<};\;
F^{+-}=F^{>};\; F^{--}=F^c;\; F^{++}=F^a$.}  reads
\begin{eqnarray}\label{KB-Eq}
&&\left(\Gr^{-1}_{0}(-\ii\partial_1)-\Gr^{-1}_{0}(-\ii\partial_2)\right)
\Gr^{-+}(1,2) \cr
&&
=\oint\di 3 \left(\Se(1^-,3 )\Gr(3 ,2^+) -\Gr(1^-,3 )\Se(3,2^+)\right)
\equiv \MP C(1^-,2^+)
\end{eqnarray}
with Fourier transform of the inverse free Green function
\begin{eqnarray}
\Gr^{-1}_{0}(p)=\left\{
\begin{array}{ll}
p^2-m^2\quad&\mbox{for relativistic bosons}\\
p_0-{\vec p}^2/(2m)\quad&\mbox{for non-rel. fermions or bosons.}
\end{array}\right.
\end{eqnarray}
Here and below the upper/lower signs refer to fermions or bosons respectively,
$\Gr_0$ and $\Gr$ correspondingly denote the free and full Green functions,
labels for the 
different species and internal quantum numbers are suppressed. The driving
term on the r.h.s. of Eq. (\ref{KB-Eq}), summarized by $C$, is a functional of
the Green 
functions through the self-energies $\Se$ contour folded with $\Gr$. The
real-time integration contour, cf. Fig. \ref{Fcontour}, is denoted by
${\cal C}$.
The step towards transport equations is provided by introducing the 
four-dimensional Wigner transforms for all two-point functions through
\begin{eqnarray}
\label{Wigner}
F(x,y)=\intp e^{-\ii p(x-y)} F({\textstyle\frac{x+y}2},p).
\end{eqnarray}
The KB Eq.  (\ref{KB-Eq}) then transforms to
\begin{eqnarray}
\label{KB-Eq-Wigner} 
v^{\mu}\partial_{\mu}(\MP\ii)\Gr^{-+}(X,p)&=& C^{-+}(X,p;\{\Gr\})
\quad\mbox{with}
\quad
v^{\mu}=\frac{\partial}{\partial p_{\mu}}G_0^{-1}(p),
\end{eqnarray}
where the r.h.s. is also expressed in terms of the Wigner transforms of all
Green functions through (\ref{Wigner}).  The final step is to expand
the complicated r.h.s. of Eq. (\ref{KB-Eq-Wigner}) to the first-order
gradients. Then the {\em local} part of this r.h.s.,
$C^{-+}_{\mathrm{(loc)}}$, consists of {\em non-gradient terms},
where one replaces the different mean positions $(x_i +x_j )/2$ occurring in
the various Green functions by the externally given mean position $X$ of the
l.h.s., i.e.  $X=(x_1+x_2)/2$ and evaluates the diagrams as in momentum
representation. The corrections for the displacement to the true coordinates
of each Green function are then accounted for to the first order in the
gradients.  Here we simply abbreviate the gradient terms by a $\Diamond$
operator acting on the {\em local} diagram expression
\begin{eqnarray}
\label{Grad-KB-Eq}
v^{\mu}\partial_{\mu}(\MP\ii)\Gr^{-+}(X,p)
&=&  (1+{\textstyle\frac{\ii}{2}}\Diamond)
\left\{C^{-+}_{\mathrm{(loc)}}(X,p)\right\},
\end{eqnarray}
where
\begin{eqnarray}
\label{C-loc}
C^{-+}_{\mathrm{(loc)}}(X,p)&=&C^{-+}(X,p;\{\Gr_{\mathrm{(loc)}}\})
,
\end{eqnarray}
is a functional of the local Green functions $\Gr_{\mathrm{(loc)}} \equiv
\Gr(X,p)$.  Here and for all further considerations below, both, Green
functions $\Gr(X,p)$ and self-energies $\Se(X,p)$, whenever quoted in their
Wigner function form, are taken in {\em local} approximation, i.e. with $X$
given by the external coordinate and $\Se(X,p)$ void of any gradient
correction terms. The explicit definition of the $\Diamond$-operator is
deferred to sect.  \ref{Gadient-Apprx}. All what we need to know at this level
is that it consists of terms where in the diagrams defining
$C_{\mathrm{(loc)}}$ pairs of Green functions are replaced by their space-time
and momentum derivatives, respectively, just leading to equations linear in
space-time gradients. Naturally the result of the $\Diamond$-operation depends
on the explicit form, i.e. diagrammatic structure, of the functional on which
it operates. Therefore for the non-gradient term in Eq. (\ref{Grad-KB-Eq}),
which defines the {\em local} collision term 
\begin{eqnarray}
\label{C-loc-dia}
C^{-+}_{\mathrm{(loc)}}
&{\begin{array}[t]{c}=\\[-1mm]
{\!\!\!\!\!\scr{diagram}\!\!\!\!\!}\end{array}}
&
\MP \Se^{-k}(X,p)\sigma_{kl}\Gr^{l+}(X,p)
-(\MP) \Gr^{-k}(X,p)\sigma_{kl}\Se^{l+}(X,p)\\
\label{C-loc-value}
&{\begin{array}[t]{c}=\\[-1mm]
{\!\!\!\scr{value}\!\!\!}\end{array}}
&
\,\underbrace{\MP \ii\Se^{-+}(X,p)\ii\Gr^{+-}(X,p)}_{\mbox{gain}}
\, -\,
\underbrace{(\MP)\ii\Gr^{-+}(X,p)\ii\Se^{+-}(X,p)}_{\mbox{loss}}
,
\end{eqnarray}
we give
both, the diagram expression (\ref{C-loc-dia}) and the normally quoted {\em
value} expression (\ref{C-loc-value}). The latter simplifies due to a
cancellation of terms which however survive for the order sensitive gradient
operation. 
Here $\sigma_{ik}=\sigma^{ik}=\mathrm{diag}(1,-1)$ defines the ``contour
metric'', which accounts for the integration sense, and summation over the
contour labels $k,l\in \{-,+\}$ is implied, cf. (\ref{sig}) ff.

The above quantum kinetic equation (\ref{Grad-KB-Eq}) has to be supplemented
by a {\em local} Dyson equation for the retarded Green function \cite{Kad62}
\begin{eqnarray}\label{retarded-Eq}
\left(\Gr^R(X,p)\right)^{-1}=\left(\Gr^R_0(p)\right)^{-1}-\Sigma^R(X,p),
\end{eqnarray}
which together with Eq. (\ref{Grad-KB-Eq}) provides the simultaneous solution
to $\Gr^{+-}$. The full retarded Green function $\Gr^R$
depends on the retarded self-energy
$\Se^R=\Se^{--}-\Se^{-+}=\Se^{+-}-\Se^{++}$ again in {\em local}
appro\-ximation.  $\Gr^R_0$ is the free retarded Green function.  Please note
that equation (\ref{retarded-Eq}) is just algebraic although it is obtained in
the framework of the first-order gradient approximation.

In most presentations of the gradient approximation to the KB equations, Eq.
(\ref{Grad-KB-Eq}) is rewritten such that the gradient terms are subdivided
into two parts, consisting of Poisson bracket terms describing drag-
and back-flow effects, on the one side, and a memory collision term
$C^{\scr{mem}}$, on the other side, in cases when the self-energy contains
internal vertices
\begin{eqnarray}\label{Grad-Sep-KB-Eq}
v^{\mu}\partial_{\mu}(\MP)\ii\Gr^{-+}(X,p) &=&
\Pbr{\Re\Se^R,\MP\ii\Gr^{-+}}+\Pbr{\MP\ii\Se^{-+},\Re\Gr^{R}}\cr
&&+ C^{-+}_{\mathrm{(mem)}}(X,p)+C^{-+}_{\mathrm{(loc)}}(X,p),
\end{eqnarray}
where
\begin{eqnarray}\label{C-mem}\nonumber
&&\hspace*{-15mm}\MP C^{-+}_{\mathrm{(mem)}}(X,p)\\
&=&\Se^{-k}_{(\scr{mem})}(X,p)\sigma_{kl}\Gr^{l+}(X,p)
-\Gr^{-k}(X,p)\sigma_{kl}
\Se_{(\scr{mem})}^{l+}(X,p)
\\
\label{C-mem-value}
&\!\!\begin{array}[t]{c}=\\[-1mm]
{\scr{value}}\end{array}\!\!&
{-\Se_{(\scr{mem})}^{-+}(X,p)\Gr^{+-}(X,p)}
+
{\Gr^{-+}(X,p)\Se_{(\scr{mem})}^{+-}(X,p)}
\end{eqnarray}
and
\begin{eqnarray}\label{S-mem}
&&\hspace*{-15mm}\Se_{(\scr{mem})}(X,p)=
{\textstyle\frac{\ii}{2}}\Diamond\left\{\Se(X,p)\right\}. 
\end{eqnarray}
One may further introduce the spectral function $A(X,p)=-2\Im \Gr^R(X,p)$
determined by the retarded equation (\ref{retarded-Eq}), as well as the
four-phase-space distribution function $f(X,p)$, by means of $\MP\ii
G^{-+}(X,p)=f(X,p) A(X,p)$, whose evolution is governed by transport Eq.
(\ref{Grad-Sep-KB-Eq}) or equivalently by (\ref{Grad-KB-Eq}). This defines a
generalized quantum transport scheme which is void of the usual quasi-particle
assumption.  The time evolution is completely determined by initial
instantaneous values of the Green functions and their gradients at each
space--time point, which means it is {\em Markovian}, 
since the memory part of the collision term is kept only up
to first-order gradient terms. Within its validity range this transport scheme
is capable to describe slow space-time evolutions of particles with broad
damping width, such as resonances, within a transport dynamics, now
necessarily formulated in the four-dimensional phase-space.

Depending on the questions raised, the above separation (\ref{Grad-Sep-KB-Eq})
may not be always useful. For the derivation of the conservation laws only a
unified treatment of both, the Poisson brackets and memory collision terms,
reveals the symmetry among these terms, which then displays the necessary
cancellation of certain contributions such that the conservation laws
emerge. Therefore, in the forthcoming considerations we will mostly refer to
the quantum kinetic equation in the form (\ref{Grad-KB-Eq}).

\section{Conservation laws}\label{CL}

Conservations of charge and energy--momentum result from taking the charge or
four--momentum weighted traces of the transport equations (\ref{Grad-KB-Eq}).
These traces include the integration over four-momentum, as well as sums over
internal quantum numbers and all types of species $a$ with charges $e_{a}$
\begin{eqnarray}
\label{Conserv-Eq}
\hspace*{-5mm}
&&\hspace*{-10mm}\partial_{\mu}\sum_a\intp\displaystyle{e_a\choose p^{\nu}}
v^{\mu}(\MP\ii)\Gr_a^{-+}(X,p)
\cr
&&=\sum_a\intp {e_a\choose p^{\nu}}
\left(1+{\textstyle\frac{\ii}{2}}\Diamond\right)
\left\{C^{--}_{a\;\mathrm{(loc)}}(X,p)\right\}
\equiv{{Q}(X)\choose {T}^{\nu}(X)}.
\end{eqnarray}
The the charge and four-momentum leaks on the r.h.s, abbreviated as $Q$ and
$T^{\nu}$, can be represented by closed diagrams, where the two end points
($x_1^-$ and $x_2^+$ in Eq.  (\ref{KB-Eq})) coalesce, i.e. $x_1=x_2=X$. In
coordinate representation the r.h.s. corresponds to contour integrals of the
type (\ref{diffrules0}) - (\ref{diffrules1}) for which entirely retarded terms
drop out. Therefore, one even can place the two end points on the same contour
side 
(respecting the fixed order for Tad-pole terms). For definiteness we have
chosen the time ordered ($-$) branch. The external point $x_1^{-}=x_2^{-}=X$
is then the reference point with respect to which the gradients are to be
evaluated. In line with causality requirements this reference point
$X=(t,{\vec x})$ is also the {\em retarded point}. This implies that in a
real-time contour representation any contribution to (\ref{Conserv-Eq}) from
internal integrations with physical times larger than $t$ drops out,
cf. \cite{Chou,Dan90}. Using the explicit form of the contour metric
$\sigma$ (cf. Eqs. (\ref{Fij}) and (\ref{H=FG})) one obtains
\begin{eqnarray} 
\label{SeGr-GrSe}
\hspace*{-3mm}
 C^{--}_{a\;\mathrm{(loc)}}(X,p)
&=&
\mp\left(\Se^a_{-k}(X,p)\Gr_a^{k-}(X,p)
-\Gr_a^{-k}(X,p)\Se^a_{k-}(X,p)\right).
\end{eqnarray}
In the proof given in sect. \ref{Phi-Conserv} we show that the local
parts of the r.h.s. of (\ref{Conserv-Eq}) as given by
\begin{eqnarray}
\label{dotQ}
{{Q}_{\mathrm{loc}}(X)\choose {T}^{\nu}_{\mathrm{loc}}(X)}
=
\sum_a\intp {e_a\choose p^{\nu}}
C^{--}_{a\mathrm{(loc)}}(X,p)
\end{eqnarray}
entirely drop out. Also the gradient terms of $Q$ cancel, while the gradients
of the $T^{\nu}$ term compile to a complete divergence \def\E{{\cal E}}
\begin{eqnarray}
\label{Conserv-Result}
{Q}(X)\equiv 0,\quad\quad 
{T}^{\nu}(X)=g^{\mu\nu}
\partial_{\mu}
\left(\E^{\mathrm{pot}}(X)-\E^{\mathrm{int}}(X)\right),
\end{eqnarray}
provided the self-energies are derived from a so called $\Phi$-functional
\cite{Baym}. This implies exact conservation
laws for the Noether currents and the energy--momentum tensor given by
\begin{eqnarray}
\label{Q-E-M}
\partial_{\mu}J^{\mu}(X)&=&0,\quad\partial_{\mu}
\Theta_{\scr{loc}}^{\mu\nu}(X)=0
\hspace*{1cm}\mbox{with}
\\
J^{\mu}(X)&=&\sum_a\intp{e_a}v^{\mu}(\MP\ii)\Gr_a^{-+}(X,p),\\
\Theta_{\scr{loc}}^{\mu\nu}(X)
&=&\sum_a\intp v^{\mu}p^{\nu}(\MP\ii)\Gr_a^{-+}(X,p) 
\cr
&&\hspace*{5mm}+g^{\mu\nu}\left(\E^{\mathrm{int}}_{\scr{loc}}(X)-
\E^{\mathrm{pot}}_{\scr{loc}}(X)\right).
\end{eqnarray}
Here $\Gr$ is the self-consistent propagator solving the coupled set of
quantum transport equations (\ref{Grad-KB-Eq}) and
(\ref{retarded-Eq}). Furthermore, $\Theta^{\mu\nu}_{\scr{loc}}(X)$ is the {\em
local} version of energy--momentum tensor which for the $\Phi$-derivable
approximation to the KB equations has been constructed in our previous papers
\cite{IKV,IKV99}. Thereby, $\E^{\mathrm{int}}_{\scr{loc}}(X)$ and
$\E^{\mathrm{pot}}_{\scr{loc}}(X)$ 
define the interaction and single-particle potential energy
densities, respectively, also taken in the local approximation. 

In the subsequent sections we formally define the complete first-order
gradient expansion for any two point function, which contains internal
vertices, and specify the corresponding diagrammatic rules. Finally, in
sect. \ref{Phi-Conserv} we prove the conservation laws (\ref{Q-E-M}),
using the $\Phi$-derivable properties and the gradient rules.

\section{Complete Gradient Approximation}\label{Gadient-Apprx}

Let $M(1,2)$ be any two-point function with complicated internal structure. We
are looking for its Wigner function $M(X,p)$ with $X=\frac{1}{2}(x_1+x_2)$ to
first-order gradient approximation. The zero-order term is just given by
evaluating $M(1,2)$ with the Wigner functions of {\em all} Green functions
taken at the same space-time point $X=(x_1+x_2)/2$ and the momentum
integrations being done as in the momentum representation of a homogeneous
system. To access the gradient terms related to any Green function $G(i,j)$
involved in $M(1,2)$, its Wigner function $G(\frac{1}{2}(x_i+x_j),p)$ is to be
Taylor expanded with respect to the reference point $X=(x_1+x_2)/2$, i.e.
\begin{eqnarray}
G(\frac{x_i+x_j}{2},p)\approx G(X,p)
+\frac{1}{2}\left[(x^{\mu}_i-x_1^{\mu})+(x_j^{\mu}-x_2^{\mu})\right]
\frac{\partial}{\partial X^{\mu}}G(X,p). 
\end{eqnarray}
Both the space derivatives of Green functions and the factors $(x_i-x_1)$
and $(x_j-x_2)$ accompanying them can be taken as special two-point functions,
and we therefore assign them special diagrams \unitlength0.75mm
\begin{eqnarray}\label{x-dashed}
\Gdashed
\;&=&\;{\textstyle\frac{1}{2}}\left(\partial _i +\partial_j\right)\Gr(i,j)
\longrightarrow \partial_X \Gr(X,p), \\
\label{p-dashed}
\Pdashed{i}{j}
\;&=&\;-\ii\left(x_i-x_j\right)
\longrightarrow -(2\pi)^4\frac{\partial}{\partial p}\delta(p) 
\end{eqnarray}
with the corresponding Wigner functions at the right hand side. Then the
gradient 
terms of a complicated two-point function (given in different notation) can
graphically be represented by the following two diagrams on the r.h.s.
\begin{eqnarray}\label{Gradient-Diag}
\Diamond \left\{M(1,2)\right\}&=&
\Diamond
\FBox{$1$}{$2$}{$M$}\!
\equiv\!\FBox{$1$}{$2$}{$\Diamond M$}
\!=\!
\FBoxL+\FBoxR
\end{eqnarray}
Here the diamond operator $\Diamond$, as above,  formally defines the gradient
approximation of the two-point function $M$ to its right with respect to 
the two
external points $(1,2)$ displayed by full dots. The diagrammatic rules are
then the 
following. For any $\Gr(3,4)$ in $M$, take the spatial derivative
$\partial_X\Gr(X,p)$ (double line) and construct the two diagrams, where
external point 1 is linked to 3 by an oriented dashed line, and where point
2 is linked to 4, respectively. Interchange of these links provides the same
result. Here $M'$ is a four-point function generated by opening $M(1,2)$
with respect to any propagator $G(3,4)$, i.e.
\begin{eqnarray}
M'(1,2;3,4)=\MP\frac{\delta M(1,2)}{\delta \;\ii G(4,3)}.
\end{eqnarray}
The diagrams in Eq. (\ref{Gradient-Diag}) are then to be evaluated in the local
approximation, i.e. with all Wigner Green functions taken
at the same space-time point $X$. The dashed line (\ref{p-dashed}) adds a new
loop integration to the diagram, which, if integrated, leads to momentum
derivatives of the Green functions involved in that loop. Both, double and
dashed lines have four-vector properties, and the rule implies a four-scalar
product between them. 

The explicit properties of the dashed line (\ref{p-dashed}) permit to
decompose it into two or several dashed lines through the algebra
\unitlength0.75mm
\begin{eqnarray}\label{p-dashed-addition}
\Pdashed{1}{3}\;&=&\;\Pdashed{1}{2}\;+\;\;\Pdashed{2}{3}
\quad\quad{\rm and}\quad\quad
\Ploop
=0.
\end{eqnarray}
In momentum-space representation these rules correspond to the partial
integration.  They imply that $\Diamond \{M(1,2)\}=0$, if $M$ contains no
internal vertices.  Applying rule (\ref{p-dashed-addition}) to the convolution
of two two-point functions
\begin{eqnarray}\label{Convolution}
C(1,2)=\oint \di 3 A(1,3)B(3,2)
\end{eqnarray}
leads to the following convolution theorem for the gradient approximation
\begin{eqnarray}\label{Poisson}
\hspace*{-5mm}\Diamond \{C(X,p)\}=&\Diamond& 
\left\{\vphantom{\int}\right.\!\DRhomb{A}{B}\!\left.\vphantom{\int}\right\}\cr
=&&
\DPoisson
\cr
&+&\DRhomb{$A$}{$\Diamond B$} +  \DRhomb{$\Diamond A$}{$B$}
\\[3mm]
=&&
\Pbr{A(X,p),B(X,p)}\cr
&&+A(X,p)\Diamond\{B(X,p)\} +\Diamond\{A(X,p)\}B(X,P).
\end{eqnarray}
Besides the standard Poisson bracket expression $\Pbr{A,B}$ it leads to
further gradients within each of the two functions (note that the $\Diamond$
operator acts only on the two-point function in the immediate braces to its
right). Applied to the r.h.s. of Eq. (\ref{Grad-KB-Eq}), this rule indeed
provides the decomposition into Poisson bracket and memory terms for the
more conventional formulation of the quantum kinetic equation
(\ref{Grad-Sep-KB-Eq}).
 
\section{$\Phi$-derivable scheme and exact conservation laws}
\label{Phi-Conserv} 

In this section we first review the $\Phi$-derivable properties at the level
of self-consistent Dyson or KB equations, then proceed towards the implications
for the quantum kinetic equations (\ref{Grad-KB-Eq}) and to the
proof of the corresponding exact conservation laws (\ref{Q-E-M}).

In a $\Phi$-derivable scheme the self-energies are generated from a functional
$\Phi\{\Gr,\lambda\}$ through the following functional variation\footnotemark,
cf.  \cite{IKV}
\begin{eqnarray}\label{varphdl1}
-\ii \Se(x,y)
&=&\mp\frac{\delta\ii \Phi\{\Gr,\lambda\}}{\delta \ii\Gr(y,x)}\times
\left\{
\begin{array}{ll}
2\quad&\mbox{for real fields}\\
1\quad&\mbox{for complex fields}
\end{array}\right. .
\end{eqnarray}
\footnotetext{Here we include also the rule for real fields, upper/lower signs
refer to fermions/bosons} The $\Phi$ functional itself is given by
two-particle irreducible ($2PI$) closed diagrams in terms of {\em full} Green
functions of the underlying field theory, i.e.
\begin{eqnarray}\label{Phi-def}
\ii\Phi\{\Gr,\lambda\}=
\left<\exp\left(\ii\oint\di^4 x \lambda(x) {\cal L}^{\rm
      int}\right)\right>_{2PI}. 
\end{eqnarray}
The interaction strength $\lambda(x)$, the physical value of which is
$\lambda=1$, allows the definition of the interaction energy density 
\begin{eqnarray}
\label{eps-int} 
{\cal E}^{\scr{int}}(x)&=&\left<-\Lint(x)\right>
=-\left.\frac{\delta\ii\Phi}{\delta\ii\lambda(x)}\right|_{\lambda=1}.
\end{eqnarray}
The single-particle potential energy density is defined as
\begin{eqnarray}
\label{eps-pot}
{\cal E}^{\scr{pot}}(x)
&=& \frac{1}{2}
\oint\di^4 y \left[
\Se(x,y) (\MP\ii)\Gr(y,x)+(\MP\ii)\Gr(x,y)\Se(y,x)\right]  
\end{eqnarray}
for complex fields. The local approximants to both ${\cal E}^{\scr{int}}$
and 
${\cal E}^{\scr{pot}}$ enter the energy--momentum tensor (\ref{Q-E-M}).

The diagrammatic series of $\Phi$ given by Eq. (\ref{Phi-def}) can be
truncated at any level. Keeping the variational property (\ref{varphdl1}), this
defines a truncated self-consistent scheme for the KB equations
(\ref{KB-Eq}). The so constructed self-energies lead to a coupling between the
different species $a$, which obey detailed
balance\footnote{generalizing the special recipes for broad resonances given
in ref.\cite{DB91}}. It has been shown \cite{Baym}, see also \cite{IKV}, that
such a self-consistent scheme is exactly conserving at the
expectation value level and thermodynamically consistent at the same time.

A prominent example is the particle-hole ring resummation in Fermi li\-quid
theory in the limit of zero range four-fermion coupling with the following diagrams
\begin{eqnarray}\label{F-liquid}
\Phi&=&{1\over 2}\PhiHartree+{1\over 4}\PhiSandwich+
\;\sum_{n>2}\frac{1}{2n}\PhiRing{4},\\
{\cal E}^{\scr int}(X)&=&
{1\over 2}\EintHartree+{1\over 2}\EintSandwich+\;\sum_{n>2}
\;{1\over 2}\;\EintRing{4},\\
\Sigma(x,y)&=&\phantom{{1\over 2}}\SigmaHartree+\phantom{{1\over 2}}
\SigmaSandwich+\;\sum_{n>2}\;\,\SigmaRing{3}
\end{eqnarray}
Here $n$ counts the number of vertices in the diagram, the full dots denote
the external points $X$ and $(x,y)$. Note that compared to ${\cal
E}^{\scr{int}}(X)$ all coordinates are integrated in $\Phi$, and therefore its
diagrams attain an extra combinatorial factor $1/n$ \cite{Luttinger}. These
diagrams illustrate various levels of approximation.  Restricting $\Phi$ just
to the one-point term provides the Hartree approximation.  From the two-point
level on the collision term becomes finite, which also leads to finite damping
widths of the particles. Terms, where $\Phi$ has more than two internal
vertices, give rise to memory contributions to the self-energies,
cf. (\ref{C-mem}), due to the intermediate times of the internal vertices in
$\Sigma(x,y)$. For this example the memory terms give rise to the famous
$T^3\ln T$ term in the specific heat of of liquid $^3$He
\cite{Riedel,Carneiro,Baym91} at low temperatures $T$.

It is possible to transcribe the variational rules to the local
approximation defining a local $\Phi$-functional replacing everywhere $\Gr$ by
its local approximant, cf. (\ref{C-loc}),
\begin{eqnarray}\label{Phi-loc}
\Phi_{\scr{loc}}(X)=\left.\Phi\{\Gr,\lambda^{\mp}\}
\right|_{\Gr=\Gr_{\scr{loc}}=\Gr(X,p)}.
\end{eqnarray}
Here $X$ is an externally given parameter defining the reference point for
the local approximation,
and $\lambda^{\mp}$
denotes the scaling factors of the vertices on the time or anti-time ordered
branches of the contour. The variational rules (\ref{varphdl1}) and
(\ref{eps-int}) then transcribe to
\begin{eqnarray}
\label{var-Phi-loc}
-\ii \Se_{ik}(X,p)
&=&\mp\frac{\delta\ii \Phi_{\scr{loc}}(X)}{\delta \ii\Gr^{ki}(X,p)}\times
\left\{
\begin{array}{ll}
2\quad&\mbox{for real fields}\\
1\quad&\mbox{for complex fields}
\end{array}\right. ,\\
\label{eps-int-loc} 
{\cal E}^{\scr{int}}_{\scr{loc}}(X)&=&
-\left.\frac{\delta\ii\Phi_{\scr{loc}}(X)}{\delta\ii\lambda^-}
\right|_{\lambda^-=1}
=\left.\frac{\delta\ii\Phi_{\scr{loc}}(X)}{\delta\ii\lambda^+}
\right|_{\lambda^+=1}.
\end{eqnarray}

In order to prove that the exact conserving properties
indeed survive in the first-order gradient expansion (\ref{Grad-KB-Eq}),
we shall use the conserving properties of the $\Phi$ diagrams at
each internal vertex together with the variational property (\ref{varphdl1}).


\subsection{Properties of diagrams}

We return to the r.h.s. terms of the conservation laws (\ref{dotQ}). Due to
the variational property (\ref{varphdl1}), it is evident that they are given
by closed diagrams of the same topology as those of $\Phi$, however, with one
point not contour integrated but placed on the time-ordered ($-$) branch with
coordinate $X$.  To further reveal this relation between the ``two-point''
representation, as given by the r.h.s. of Eq. (\ref{dotQ}), and the {\em
retarded} one-point representation of $\Phi$ with respect to a chosen
reference point $X^-$, we have to resolve the internal structure of the
different diagrams contributing to $\Phi$.  Therefore, we first decompose
$\Phi$ into the terms of different diagrammatic topology
\begin{eqnarray}\label{Phi_D}
\ii\Phi_{\scr{loc}}(X)=\sum_{D}\ii\Phi_{\scr{loc}}^{D}(X)
\end{eqnarray}
and discuss the features of any such term $\Phi_{\scr{loc}}^D$.  We then
enumerate the different vertices ($r=1,\dots,n_\lambda$) in each
$\Phi_{\scr{loc}}^D$, $n_\lambda$ denoting the number of vertices in
$\Phi_{\scr{loc}}^D$. For each such retarded vertex $r^-$ we define a
one-point function $\Phi_{\scr{loc}}^D(X;r^-)$, which is obtained by
integration and $\{-+\}$ summation over all other vertices except $r$ and by
putting $r$ on the time ordered ($-$) branch at coordinate $X$.  Subsequently
we enumerate the few Green functions attached to $r^-$ and label them as
$\Gr^{-k}_\gamma$ or $\Gr^{k-}_{\bar{\gamma}}$, depending on the line sense
pointing towards or away from $r^-$, respectively. Thus, the retarded function
$\Phi^D(X;r^-)$ can be represented by a diagram, where the retarded point is
explicitly pulled out. We draw this as a kind of ``parachute'' diagram which in
{\em local} approximation becomes
\begin{eqnarray}
\label{parachute}
\hspace*{-10mm}
\ii\Phi_{\scr{loc}}^D(X;r^-)&=&
\unitlength1.0mm
\Parachute
\cr
&=& \int\dpi{p_1}\cdots\dpi{\bar{p}_1}\cdots
{\cal C}^{Dr}_{k_1\cdots \bar{k}_1\cdots}
(X;p_1,\dots,\bar{p}_1,\dots)\cr
&&\hspace*{10mm}\times\;(\mp\ii) G^{-k_1}_1(X,p_1)\cdots 
\ii G^{\bar{k}_1-}_{\bar{1}}(X,\bar{p}_1)\cdots  
\end{eqnarray}
The Green functions attached to the external retarded point $r^-$, form the
``suspension cords''.  The ``canopy'' part ${\cal C}^{Dr}$ specifies the rest
of the diagram which is not explicitly drawn\footnote{For definiteness we have
drawn a diagram with 4 propagators linking to the retarded point $r^-$. All
considerations, however, are independent of the coupling scheme which can even
vary from vertex to vertex. For simple $\Phi$ diagrams, such as those with
only two vertices at all, the canopy part reduces simply to a single point,
cf. the second diagram in Eq. (\ref{F-liquid}).}.
Here all Green functions are taken in the local approximation, i.e. with
coordinate $X$ given by the reference point.  From the variational principle
(\ref{varphdl1}) it is clear that $\Phi^D(X;r^-)$ can be interpreted in
different equivalent ways depending on which line attached to $r^-$ being
opened \unitlength1mm
\begin{eqnarray}
\label{parachute1}
\hspace*{-8mm}
\ii\Phi_{\scr{loc}}^D(X;r^-)&=&\MP
\int\frac{\di^4 p}{(2\pi)^4}\Gr^{-k}_\gamma(X,p)
\Se^{Dr\gamma}_{k-}(X,p)
\\
\label{parachute2}
&=&\MP\int\frac{\di^4 p}{(2\pi)^4}
\Se^{Dr\bar{\gamma}}_{-k}(X,p)
\Gr^{k-}_{\bar{\gamma}}(X,p) 
,
\end{eqnarray}
Here $\Gr_\gamma$ (or $\Gr_{\bar{\gamma}}$) is one of the suspension cords
with the arrow pointing towards (or away) from $r$. {\em No} summation over
$\gamma$ is implied by these relations! The self-energy terms $\Se^{Dr\gamma}$
and $\Se^{Dr\bar{\gamma}}$ are given by those subdiagrams of
$\Phi_{\scr{loc}}^{D}(r^-)$ complementary to $\Gr_\gamma$ and
$\Gr_{\bar{\gamma}}$, respectively.  According to the variational rule
(\ref{varphdl1}) they contribute to the self-energy of a given species $a$
as
\begin{eqnarray}
\label{Dl-sum}
-\ii\Se^a_{k-}(X,p) &=& \MP\frac{\delta\ii\Phi_{\scr{loc}}}{\delta
\ii\Gr_a^{-k}(X,p)}
= 
-\sum_{D}\sum_{r\in D}\sum_{\gamma}\ii\Se^{Dr\gamma}_{k-}(X,p)\delta_{a\gamma}
\end{eqnarray}
and similarly for $\Se^{Dr\bar{\gamma}}_{-k}$ leading to $\Se^a_{-k}$. The
Kronecker symbol $\delta_{a\gamma}$ projects on species $a$. The summation in
$r$ runs over all vertices in the $\Phi_{\scr{loc}}^{D}$ diagram.  It is clear
that each Green function in $\Phi_{\scr{loc}}^D$ appears once in the count of
$\Se_{k-}$ and once in the count of $\Se_{-k}$. This precisely matches with
the two terms required in Eq.  (\ref{SeGr-GrSe}). We further note that the
variation (\ref{eps-int}) of $\Phi$ with respect to the coupling strength
$\lambda(X)$ can be represented as
\begin{eqnarray}\label{var-lambda}
{\cal E}^{\scr{int}}(X)
=-\frac{\delta\ii\Phi_{\scr{loc}}(X)}{\delta \ii\lambda^-}
&=&
-\sum_{D}\sum_{r\in D}\Phi_{\scr{loc}}^D(X;r^-).
\end{eqnarray}

\subsection{Charge-current conservation}

The discussion above shows that any term of the local part of the charge leak
$Q_{\mathrm{loc}}(X)$ in Eq. (\ref{dotQ}) is given by a closed diagram of
the same topology as the $\Phi_{\scr{loc}}^{D}(r^-)$. Indeed, weighting
$\Phi_{\scr{loc}}^D(X;r^-)$ with the charges of the in- and out-going Green
functions reproduces piece by piece the local gain and loss terms in Eq.
(\ref{SeGr-GrSe}). The sum over all possible retarded vertices $r$ in diagram
$D$ and the sum over all diagrams indeed exactly construct
$Q_{\scr{loc}}$ as
\begin{eqnarray}\label{Q.loc-1}
\hspace*{-4mm}Q_{\scr{loc}}(X)&=&
\intp\sum_a e_a C^{--}_{a,\;\mathrm{loc}}(X,p)
\cr
&=&
\sum_{D}\sum_{r\in D}
\left(\vphantom{\sum_i}\right.
\underbrace{
\sum_{\bar{\gamma}} e_{\bar{\gamma}}-\sum_\gamma e_{\gamma}
}_{\equiv 0}
\left.\vphantom{\sum_i}\right)
\ii\Phi_{\scr{loc}}^D(X;r^-).
\end{eqnarray}
Since the charge sum vanishes at each vertex, $Q_{\scr{loc}}$
identically vanishes.

The fact that the local collision term part vanishes is indeed trivial. The
point here is that the cancellation occurs diagram by diagram in terms of
$\Phi_{\scr{loc}}^D(X;r^-)$. This has the important consequence that the
gradient terms exactly cancel out, too, i.e.
\begin{eqnarray}
\label{gradQ=0}
\sum_a\intp e_a
\Diamond C^{--}_{a,\;\mathrm{loc}}(X,p)
= 
\Diamond Q_{\scr{loc}}\equiv 0, 
\end{eqnarray}
since they are generated by applying linear differential operations to the
integrand of $\Phi_{\scr{loc}}^D(X;r^-)$, while the charge factors are
constants. Therefore, we have verified that $Q(X)\equiv 0$. This proves
current conservation, where the current has the original Noether form. This
proof applies to any conserved current of the underlying field theory, which
relates to a global symmetry.

\subsection{Energy-Momentum Tensor}

In a similar way as above the four-momentum weighted
terms become
\begin{eqnarray}\label{T.2}
\hspace*{-0.5cm}
&&T^{\nu}_{\mathrm{loc}}(X)=\sum_{D}\sum_{r\in D}
\int\dpi{p_1}\cdots\dpi{\bar{p}_1}\cdots
\left(\vphantom{\sum_i}\right.
\underbrace{
\sum_{\bar{\gamma}} p^{\nu}_{\bar{\gamma}}-\sum_\gamma p^{\nu}_{\gamma}
}_{\equiv 0}
\left.\vphantom{\sum_i}\right)
\\\nonumber
&&\times
{\cal C}^{Dr}_{k_1\cdots \bar{k}_1\cdots}
(X;p_1,\dots,\bar{p}_1,\dots)
(\mp\ii) G^{-k_1}_1(X,p_1)\cdots 
\ii G^{\bar{k}_1-}_{\bar{1}}(X,\bar{p}_1)\cdots . 
\end{eqnarray}
Here we have used the integrand form (\ref{parachute}) of the parachute
diagram $\Phi(X;r^-)$, since now the weights are momentum dependent.  This
local collision-term part again drops out in the same way as above. However,
the gradient correction to ${T}^{\nu}$ now involves momentum derivatives which
can act on the $p^{\nu}$-factors in Eq. (\ref{Conserv-Eq}) through partial
integrations. Indeed, {\em only} those momentum derivatives survive that act
on any of the $p^{\nu}$-factors, since momentum derivatives acting on any of
the Green functions leave the vanishing pre-factor in Eq. (\ref{T.2})
untouched.

The actual evaluation of the gradients is subtle and depends on the detailed
topological structure of the diagram.  For this purpose we use the
diagrammatic rules (\ref{Gradient-Diag}) for the gradient terms with the
double line as $\partial^{\mu}G(X,p)$, dashed lines as
$\propto\partial\delta(p)/\partial p^{\mu}$. For the subsequent analysis we
also introduce a new diagrammatic element, a genuine two-point function which
gives the four-momentum $p^\nu$-factor  ``flowing'' through the line
\begin{eqnarray}\label{p-graf}
p^\nu=\Pnu
\quad\quad\mbox{with}\quad
\PnuLr
= -\PnuLl
= \frac{\partial p^\nu}{\partial p_\mu}=g^{\mu\nu}. 
\end{eqnarray}
The last relation results from partial integration and depends only on the
sense of the dashed line.  

Note that the $p^\nu$ factors in the r.h.s. of Eq. (\ref{Conserv-Eq}) occur
outside the gradient expression.  Using the convolution theorem
(\ref{Poisson}), we first pull the $p^\nu$ factors into the gradient terms in
the following manner (using $g^{\mu\nu}={\partial p^\nu}/{\partial p_\mu}$)
\begin{eqnarray}\label{convol-pnu}
&&\Diamond\left\{\Se\cdot\Gr\right\}p^\nu 
-p^\nu\Diamond\left\{\Gr\cdot\Se\right\}
\cr
&&\hspace*{5mm} =
\underbrace{g^{\mu\nu}
\partial_{\mu}
\left(\Se\cdot\Gr+\Gr\cdot\Se \right)}_{\MP E_1^{\nu}(X,p)}
+
\underbrace{
\Diamond
\left\{\Se\cdot\Gr\cdot p^\nu-p^\nu\cdot\Gr\cdot\Se \right\}}_{\MP E_2^{\nu}
(X,p)}.
\end{eqnarray}
Please, note the order of the terms under the $\Diamond$ operator as they do
{\em not commute}! The first term, $E_1^\nu$, arose from the Poisson bracket
term in Eq. (\ref{Poisson}). In accord with Eq. (\ref{eps-pot}), the
four-momentum integration obviously provides the 
potential energy density term 
\begin{eqnarray}\label{E-1}
{\textstyle\frac{\ii}{2}}\intp E_1^{\nu}(X,p)=g^{\mu\nu}
\partial_\mu  {\cal E}^{\scr{pot}}_{\scr{loc}}(X)
\end{eqnarray}
of the energy--momentum tensor (\ref{Q-E-M}). 

Thus, we expect the second $E_2^\nu$ term to generate the remaining
interaction energy density part $g^{\mu\nu}{\cal
E}^{\scr{int}}_{\scr{loc}}(X)$ of $\Theta^{\mu\nu}$.  The four-momentum
integration of the $E_2^\nu$ term again leads to a coalescence of the two
external points. Thus the corresponding diagrams are of the parachute type
\unitlength0.75mm
\begin{eqnarray}\label{Grad-p}
\intp E_2^\nu(X,p)=
\sum_{D,r\in D}
\left(
\Diamond
\ParachL
+\dots +
\Diamond
\ParachR
\right) ,
\end{eqnarray}
where both reference points for the gradient expansion coalesce to $r^-$,
cf. Eqs. (\ref{parachute1}) and (\ref{parachute2}). Here the $p^\nu$ factor
occurs in sequence at each of the suspension cords reflecting the sum over
$\gamma$ and $\bar{\gamma}$. In order to exploit the fact that thereby
$\sum_\gamma p^\nu_\gamma-\sum_{\bar{\gamma}} p^\nu_{\bar{\gamma}}$ vanishes,
all the different diagrams in the bracket in (\ref{Grad-p}) have to be
evaluated in the same way. Omitting all labels, one obtains for the first
diagram in Eq. (\ref{Grad-p})
\begin{eqnarray}\label{grad-p-para}
\hspace*{-7mm}\Diamond
\ParachAll&=&\Paracha+\Parachb_{\vphantom{\int}}\cr
&&+\Parachc+\Parachd+\dots+\Parachf_{\vphantom{\int}}\\
\label{Par-survive}
&\Longrightarrow&2\;\Parachalpha\;+\; 2\;\Parachbeta
\end{eqnarray}
where ${{\cal C}'}=\MP\delta {\cal C}/\delta\ii\Gr$ and the four-scalar
product between the double and dashed lines is implied in each diagram. The
sequence (a) to (f) defines all diagrams resulting from the gradient
expansion. The two diagrams ($\alpha$) and ($\beta$) shown in
(\ref{Par-survive}) are the only ones that finally survive the $p^{\nu}$ sum,
i.e. the sum over ``suspension cords'' in Eq. 
(\ref{Grad-p}). In detail: diagrams (a) and (b) specify the gradient
terms arising from the space-time derivatives acting on Green functions within
the canopy part. Using addition theorem (\ref{p-dashed-addition}) the dashed
lines can be first linked from the bottom point to one definitely chosen upper
suspension points {\bf f}, as shown in diagram ($\alpha$), and from there then
further linked to the end points of the double line.  However, the latter
terms lead to $p$-derivatives entirely within the canopy, which finally drop
out due to the vanishing $p^\nu$ sum. Diagrams (c) to (f) sequentially take
the space-time gradients of the Green functions in the suspension cords
converting them to a double line in each case. Also here only terms survive,
where the momentum derivative acts on the $p^\nu$ factor, leading to digram
($\beta$).

One can now compile the terms in Eq. (\ref{Grad-p}) provided point {\bf f} is
kept fixed. Then from ($\alpha$) only a single term survives, namely that
where the $p^\nu$ factor is on the line linking to {\bf f}, while from
($\beta$) each term survives. The net result leads to a common
$g^{\mu\nu}$-factor, cf. (\ref{p-graf}), times a total derivative of the
entire parachute diagram, i.e.
\begin{eqnarray}\label{Grad-p-new}
{\textstyle\frac{\ii}{2}}\intp E_2^\nu(X,p)&=&
-\ii\sum_{D,r\in D}
g^{\mu\nu}\partial_{\mu}
\ParachEint
\cr
&=&-\ii g^{\mu\nu}\partial_{\mu}
\underbrace{\sum_{D,r\in D}\ii\Phi_{\scr{loc}}^D(X,r^-)}
_{\displaystyle
\ii\frac{\delta\ii\Phi_{\scr{loc}}^D}{\delta\ii\lambda^-}}\\
&=&-g^{\mu\nu}\partial_{\mu}{\cal E}^{\scr{int}}(X)
\end{eqnarray}
from Eq. (\ref{var-lambda}). Together with relation (\ref{E-1}), this gives
\begin{eqnarray}
\label{R(p)-Diag2}
T^{\nu}(X)=
-\partial^{\nu} 
\left({\cal E}^{\rm int}_{\rm loc}(X)-{\cal E}^{\rm pot}_{\rm loc}(X)\right)
\end{eqnarray}
for the r.h.s of conservation law (\ref{Conserv-Eq}). It is seen to be
determined by the full divergence of the difference between the interaction
energy and single-particle potential energy densities defined by
Eqs. (\ref{eps-int}) and (\ref{eps-pot}), now however evaluated
in the {\em local} approximation, i.e. with no gradient terms in the Wigner
representation:
\begin{eqnarray}
\label{eps-pot-local}
{\cal E}^{\scr{pot}}_{\rm loc}(X)
&=&
\int\dpi{p} \left[
\Re\Sa^R(X,p)(\MP\ii)\Gr^{-+}(X,p)\right.\cr
&&\hspace*{17mm}+\left.\Re\Gr^R(X,p)(\MP\ii)\Sa^{-+}(X,p)
\right]\\
&=&-
\int\dpi{p} \left[
\Re\left(\frac{\delta\Phi_{\scr{loc}}}{\delta\ii\Gr(X,p)}\right)^R
\ii\Gr^{-+}(X,p)\right.\cr
&&\hspace*{19mm}+\left.\Re\Gr^{R}(X,p)\left(\frac{\delta\Phi_{\scr{loc}}}
{\delta\ii\Gr(X,p)}\right)^{-+}
\right].
\end{eqnarray}
Here the superscript $R$ denotes the corresponding retarded function.  For
each diagram $\Phi_{\scr{loc}}^D$ the contribution to ${\cal
E}^{\scr{int}}_{\rm loc}(X)$ results from the corresponding terms of ${\cal
E}^{\scr{pot}}_{\rm loc}(X)$ just by scaling each term by the number of
vertices $n_\lambda$ over the number of Green functions $n_G$.

\section{Concluding remarks}
The quantum transport equations in the form originally proposed by Kadanoff
and Baym, (\ref{Grad-KB-Eq}) or equivalently (\ref{Grad-Sep-KB-Eq}), have very
pleasant generic features. As possible memory effects in the collision term
are to be included only up to first-order space--time gradients, 
they are local in time to the extent that only the
knowledge of the Green functions and their space-time and four-momentum
derivatives at this time is required to determine the future evolution. They
further preserve the retarded relations among the various real-time components
of the Green function. 

In this paper we have shown that they also possess {\em exact} rather than
approximate conservation laws, related to global symmetries of the system, if
the scheme is $\Phi$-derivable.  The same Noether currents and the same
energy-momentum-tensor \cite{IKV} as those for the original KB equations,
however now in their local approximation forms, are exactly conserved for the
complete gradient-expanded KB equation, i.e. the quantum transport equations
(\ref{Grad-KB-Eq}). Thus,
\begin{eqnarray}\label{C-j-mu}
\partial_{\mu}J^{\mu}(X)&=&0,\quad\quad
\label{C-Th-mu-nu}
\partial_{\mu}\Theta_{\scr{loc}}^{\mu\nu}(X)=0,
\quad\quad\mbox{with}\\
\label{j-mu}
J^{\mu}(X)&=&\sum_a e_a\int\dpi{p}p^{\mu}f_a(X,p)A_a(X,p),
\\
\label{Th-mu-nu}
\Theta_{\scr{loc}}^{\mu\nu}(X)&=&\sum_a 
\underbrace{\int\dpi{p}v^{\mu}p^{\nu}f_a(X,p)A_a(X,p)}
_{\mbox{\footnotesize sum of single particle terms}}\cr
&&\hspace*{30mm}
+g^{\mu\nu}
\left({\cal E}^{\rm int}_{\rm loc}(X)-{\cal E}^{\rm pot}_{\rm loc}(X)\right),
\end{eqnarray}
here written in terms of the product of phase-space occupation and spectral
functions $f_a(X,p)A_a(X,p)=(\MP\ii)\Gr^{-+}_a(X,p)$, are exact consequences
of the equations of motion. In order to preserve this exact conserving
property, two conditions have to be met. First, the original KB equations
should be based on a $\Phi$-derivable approximation scheme that guarantees
that the KB equations themselves are conserving \cite{Baym,IKV99}. The second
condition is that the gradient expansion has to be done systematically,
whereby it is important that no further approximations are applied that
violate the balance between the different first-order gradient terms.

Indeed, all gradient terms residing in the Poisson brackets and in the memory
collision term cancel each other for the conserved currents such that the
original Noether expression (\ref{j-mu}) remains conserved. This implies the
compensation of drag-flow terms by all the other gradient terms (back-flow and
memory flow), cf. the discussion given in ref. \cite{IKV99}. For the
energy--momentum tensor the gradient terms result into the divergence of the
difference between interaction energy density and single-particle potential
energy density, cf. (\ref{Th-mu-nu}).  Thereby ${\cal E}^{\rm int}_{\rm
loc}(X)$ and ${\cal E}^{\rm pot}_{\rm loc}(X)$ are obtained from the same
$\Phi$-functional in the local approximation as the self-energies driving the
equations of motion (\ref{Grad-KB-Eq}) and (\ref{retarded-Eq}). The so
obtained energy--momentum tensor is general and applies to any local coupling
scheme. The energy component $\Theta_{\scr{loc}}^{00}$ has a simple
interpretation. The first term determines the single-particle energy, which
consists of the kinetic and single-particle potential energy parts. Evidently
this part by itself is not conserved\footnote{Contrary to the constructions
given in ref.  \cite{Cass00}.}. Rather its potential energy part is
compensated by the last term, i.e. ${\cal E}^{\rm pot}_{\rm loc}(X)$, such
that finally the total kinetic plus interaction energy survive.

A typical example for the imbalance of gradient terms is the case, where one
neglects the second Poisson bracket term in (\ref{Grad-Sep-KB-Eq}). This
implies that drag-flow effects contained in the first Poisson bracket remain
uncompensated. Also possible memory effects $C_{\scr{mem}}$ in the collision
term should not be omitted. Otherwise 
the Poisson brackets
$\Pbr{\Re\Se_a^R,\ii\Gr_a^{-+}} + \Pbr{\ii\Se_a^{-+},\Re\Gr_a^{R}}$ remain
uncompensated and, as a consequence, the conservation laws are again violated
already in zero-order gradients.

A less evident example is the modification of the gradient terms after the
formal gradient expansion, as it has been suggested by Botermans and Malfliet
\cite{Bot90}, see also \cite{IKV99}. There one simplifies those self-energy
terms that are involved in the Poisson brackets
$\Pbr{\Re\Se_a^R,\ii\Gr_a^{-+}}+\Pbr{\ii\Se_a^{-+},\Re\Gr_a^{R}}$, employing
quasi-equilibrium relations. This modification implies deviations at
second-order gradients only\footnote{This freedom of choice is due to the fact
that various redundant combinations of the KB equations lead to non-redundant
equations after gradient approximation due to the asymmetric treatment of sums
and differences of the KB equations and their adjoint ones in the gradient
approximation. The so called mass-shell equation indeed agrees in the
first-order gradient terms with the here considered transport
Eq. (\ref{Grad-KB-Eq}), cf. \cite{IKV99}, however they differ in higher orders
of gradients.}, which is quite acceptable from the formal point of view, when
one considers a slow space-time dynamics. However, such kind of modifications
violate the strict balance between the gradient terms and thus lead to
approximate conservation laws though within first-order gradients.  Although
even in the BM case an exact conservation law can be formulated for an
effective charge \cite{Leupold}, which however only approximately coincides
with the true (Noether) one, exact conservation laws of energy and momentum
could not be derived yet.

The presence of exact conservations puts the Kadanoff--Baym formulation of
quantum transport to the level of a generic phenomenological concept. It
permits to define phenomenological models for the dynamical description of
particles with broad damping widths, such as resonances, with built-in
consistency and exact conservation laws, which for practical simulations of
complex dynamical systems may even be applied in cases, where the smallness of
the gradients can not always be guaranteed. This opens applications to the
strong non-equilibrium dynamics of high-energy nuclear collisions.

We considered here systems of relativistic bosons and/or non-relativistic
particles with a local interaction. The generalization to relativistic
fermions with local interactions is straightforward but involves extra
complications resulting from the spinor structure of the kinetic equations.

Non-relativistic systems with instantaneous interaction at finite spatial
distance (two-body potentials) also possess exact conservation laws, if
given by a $\Phi$-derivable approximation. Such systems can equivalently be
described by a local field theory with mesons mediating the interactions. The
appropriate limit towards a potential picture is obtained by treating these
interactions instantaneously (non-relativistic limit), i.e. without
retardation. This amounts to reduce the corresponding meson Dyson equation to
a Poisson equation over the sources of the meson fields and finally
eliminating those meson fields. As the local field theory is conserving, the
corresponding non-relativistic picture with potentials is conserving
too. Note, however, that in this latter case the spatially local
energy--momentum tensor does not exist, since energies and momenta are
transfered at finite distances through the potentials and it is only possible
to formulate the conservation of the {\em total} space-integrated energy and
momentum. 

An important example of such a conserving approximation is given by the ring
diagrams
\begin{eqnarray}\label{ring}
\ii\Phi=&\frac{1}{2}\PhiHartreeT&+\sum_{n=2}^{\infty}\frac{1}{2n}
\PhiRingT{8},\\[3mm]\label{Sigma-ring}
-\ii\Sigma(x,y)=&\;\;\SigmaHartreeT &+ \;\;\sum_{n=2}^{\infty}\;\SigmaRingT{6}, 
\end{eqnarray}
which is just the finite range analog of the Fermi-liquid example
(\ref{F-liquid}), now for the particle-particle resummation channel.  Here $n$
counts the number of interaction (dashed) lines representing the two-body
potential $V(x_i-x_k)$. The first terms in both expressions are the usual
Hartree terms. The remaining sum leads to a conserving $T$-matrix type of
approximation for the self-energies (\ref{Sigma-ring}), which, e.g., provides
thermodynamically consistent description of the nuclear matter
\cite{Bozek,Dickhoff,Dewulf}. In the dilute limit it
expresses the self-energies through the vacuum scattering
$T$-matrix\cite{Lenz,Dover71,IKV}. In this limit it provides a collision term
given by vacuum scattering cross sections and at the same time the gradient
terms account for the appropriate virial corrections.  The latter indirectly
depend on the energy variations of the corresponding phase shifts, which give
rise to delay time effects \cite{DP,IKV} and the corresponding changes of the
underlying equation of state (energy--momentum tensor)
\cite{BethU,DMB,Mekjian,VPrakash}. Furthermore, the $s$-channel bosonization
of the interactions in the particle-hole channel, cf. ref. \cite{IKV99},
leads to the RPA-approximation. Further applications and considerations of the
quantum transport equations will be discussed in a forthcoming paper.

A case that still requires a separate treatment is that of derivative
coupling of relativistic fields, as e.g., in the case of the
pion--nucleon interaction. The reason is that derivative couplings produce
extra terms in the currents and the energy--momentum tensor, which require
special treatment.

Besides all this success at the one-particle expectation value level, one has
to keep in mind that partial Dyson resummations still may violate the
symmetries at the two-body correlator level and beyond. In particular, it
means that the corresponding Ward-Takahashi identities are not necessarily
fulfilled within the $\Phi$-derivable Dyson resummation scheme.  On the
other side the $\Phi$-derivable scheme provides the tools to construct the
driving terms and kernels of the corresponding higher order vertex equations
(Bethe--Salpeter equations, etc.) which precisely recover the conservation
laws at the correlator level \cite{KadB,Baym,Hees-Thesis,HK01}. So far such
equations, however, could mostly be solved in drastically simplified cases
(e.g., by RPA-type resummation). A particular challenge represents the
inclusion of vector or gauge bosons into a self-consistent Dyson scheme beyond
the mean-field level, i.e. at the propagator level, since partial Dyson
resummations violate the four-dimensional transversality of the propagators. A
practical way out of this difficulty has recently been suggested in
refs. \cite{HK00,Hees-Thesis}. A further virtue of the $\Phi$-derivable scheme
is that it apparently permits a renormalization of the non-perturbative
self-consistent self-energies with temperature- and density-independent
counter terms \cite{Hees-Thesis}.

\section*{Acknowledgments}
We are grateful to G. Baym, P. Danielewicz, H. Feldmeier, B. Friman, H. van
Hees, C. Greiner, E.E. Kolomeitsev and S. Leupold for fruitful discussions on
various aspects of this research. Two of us (Y.B.I. and D.N.V.) highly
appreciate the hospitality and support rendered to us at Gesellschaft f\"ur
Schwerionenforschung.  This work has been supported in part by DFG (project
436 Rus 113/558/0).  Y.B.I and D.N.V. were partially supported by RFBR grant
NNIO-00-02-04012.  Y.B.I. was also partially supported by RFBR grant
00-15-96590.

\appendix{A}
\section{Contour Matrix Notation} \label{Contour}

In calculations that apply the Wigner transformations, it is necessary to
decompose the full contour into its two branches---the {\em time-ordered} and
{\em anti-time-ordered} branches. One then has to distinguish between the
phy\-sical space-time coordinates $x,\dots$ and the corresponding contour
coordinates $x^{\cal C}$ which for a given $x$ take two values
$x^-=(x^-_{\mu})$ and $x^+=(x^+_{\mu})$ ($\mu\in\{0,1,2,3\}$) on the two
branches of the contour (see figure 1).  Closed real-time contour integrations
can then be decomposed as
%
\begin{eqnarray}
\label{C-int}
\oint\di x^{\cal C} \dots &=&\int_{t_0}^{\infty}\di x^-\dots
+\int^{t_0}_{\infty}\di x^+\dots
\cr
&=&\int_{t_0}^{\infty}\di x^-\dots -\int_{t_0}^{\infty}\di x^+\dots, 
\end{eqnarray}
%
where only the time limits are explicitly given.  The extra minus sign of the
anti-time-ordered branch can conveniently be formulated by a $\{-+\}$
``metric'' with the metric tensor in $\{-+\}$ indices
%
\begin{eqnarray}
\label{sig}
\left(\sigma^{ij}\right)&=&
\left(\sigma_{ij}\vphantom{\sigma^{ij}}\right)=
{\footnotesize\left(\begin{array}{cc}1&0\\ 
0& -1\end{array}\right)}
\end{eqnarray}
%
which provides a proper matrix algebra for multi-point functions on the
contour with ``co''- and ``contra''-contour values.  Thus, for any two-point
function $F$, the contour values are defined as
%
\begin{eqnarray}\label{Fij}
F^{ij}(x,y)&:=&F(x^i,y^j), \quad i,j\in\{-,+\},\quad\mbox{with}\cr
F_i^{~j}(x,y)&:=&\sigma_{ik}F^{kj}(x,y),\quad
F^i_{~j}(x,y):=F^{ik}(x,y)\sigma_{ki}\cr
F_{ij}(x,y)&:=&\sigma_{ik}\sigma_{jl}F^{kl}(x,y),
\quad\sigma_i^k=\delta_{ik}
\end{eqnarray}
%
on the different branches of the contour. Here summation over repeated indices
is implied. Then contour folding of contour two-point functions, e.g. in Dyson
equations, simply becomes
%
\begin{eqnarray}\label{H=FG}
H(x^i,y^k)=H^{ik}(x,y)&=&\oint\di z^{\cal C} F(x^i,z^{\cal C})G(z^{\cal C},y^k)
\cr
&=&\int\di z F^i_{~j}(x,z)G^{jk}(z,y)
\end{eqnarray}
%
in the matrix notation.

For any multi-point function the external point $x_{max}$, which has the
largest physical time, can be placed on either branch of the contour without
changing the value, since the contour-time evolution from $x_{max}^-$ to
$x_{max}^+$ provides unity. Therefore, one-point functions have the same value
on both sides of the contour.

Due to the change of operator ordering, genuine multi-point functions are, in
general, discontinuous, when two contour coordinates become identical. In
particular, two-point functions like $\ii F(x,y)=\left<\Tc {\widehat
    A(x)}\medhat{B}(y)\right>$ become
%
\begin{eqnarray}\label{Fxy}
\hspace*{-0.5cm}\ii F(x,y) &=&
\left(\begin{array}{ccc} 
\ii F^{--}(x,y)&&\ii F^{-+}(x,y)\\[3mm]
\ii F^{+-}(x,y)&&\ii F^{++}(x,y)
\end{array}\right)
\cr
&=&
\left(\begin{array}{ccc} 
\left<{\cal T}\medhat{A}(x)\medhat{B}(y)\right>&\hspace*{5mm}&
\mp \left<\medhat{B}(y)\medhat{A}(x)\right>\\[5mm]
\left<\medhat{A}(x)\medhat{B}(y)\right>
&&\left<{\cal T}^{-1}\medhat{A}(x)\medhat{B}(y)\right>
\end{array}\right), 
\end{eqnarray}
%
where ${\cal T}$ and ${\cal T}^{-1}$ are the usual time and anti-time ordering
operators.  Since there are altogether only two possible orderings of the two
operators, in fact given by the Wightman functions $F^{-+}$ and $F^{+-}$,
which are both continuous, not all four components of $F$ are independent. Eq.
(\ref{Fxy}) implies the following relations between non-equilibrium and usual
retarded and advanced functions
%
\begin{eqnarray}\label{Fretarded}
F^R(x,y)&:=&\Theta(x_0-y_0)\left(F^{+-}(x,y)-F^{-+}(x,y)\right),\nonumber\\
&=&F^{--}(x,y)-F^{-+}(x,y)=F^{+-}(x,y)-F^{++}(x,y)\nonumber\\
F^A (x,y)&:=&-\Theta(y_0-x_0)\left(F^{+-}(x,y)-F^{-+}(x,y)\right)\nonumber\\
&=&F^{--}(x,y)-F^{+-}(x,y)=F^{-+}(x,y)-F^{++}(x,y),
\end{eqnarray}
%
where $\Theta(x_0-y_0)$ is the step function of the time difference.  The
rules for the co-contour functions $F_{--}$ etc. follow from Eq. (\ref{Fij}).

Discontinuities of a two-point function may cause problems for
differentiations, in particular, since they often occur simultaneously in
products of two or more two-point functions. The proper procedure is, first,
with the help of Eq.  (\ref{Fretarded}) to represent the discontinuous parts
in $F^{--}$ and $F^{++}$ by the continuous $F^{-+}$ and $F^{+-}$ times
$\Theta$-functions, then to combine all discontinuities, e.g. with respect to
$x_0-y_0$, into a single term proportional to $\Theta(x_0-y_0)$, and finally
to apply the differentiations.  One can easily check that in the
following particularly relevant cases
%
\begin{eqnarray}
\label{diffrules0}
&&
\oint\di z\left(
F(x^i,z)G(z,x^j) - G(x^i,z)F(z,x^j)\right), 
\\ 
\label{diffrules}
&&
\frac{\partial}{\partial x_{\mu}}
\oint\di z\left(
F(x^i,z)G(z,x^j)+G(x^i,z)F(z,x^j)\right),
\\ \label{diffrules1}
&&\left[\left(\frac{\partial}{\partial x_{\mu}}
              -\frac{\partial}{\partial y_{\mu}}\right)
\oint\di z \vphantom{\frac{\partial}{\partial x_{\mu}}
              -\frac{\partial}{\partial y_{\mu}}}
\left(F(x^i,z)G(z,y^j)-G(x^i,z)F(z,y^j)\right)\right]_{x=y}
\end{eqnarray}
%
{\em all discontinuities exactly cancel}. Thereby, these values are
independent of the placement of $x^i$ and $x^j$ on the contour, i.e. the
values are only functions of the physical coordinate $x$.

For such two point functions complex conjugation implies
%
\begin{eqnarray}\label{ComplexConjugate}
\left(\ii F^{-+}(x,y)\right)^*&=&\ii F^{-+}(y,x)
\quad\Rightarrow\quad \ii F^{-+}(X,p)=\mbox{real},\nonumber\\
\left(\ii F^{+-}(x,y)\right)^*&=&\ii F^{+-}(y,x)
\quad\Rightarrow\quad \ii F^{+-}(X,p)=\mbox{real},\nonumber\\
\left(\ii F^{--}(x,y)\right)^*&=&\ii F^{++}(y,x)
\quad\Rightarrow\quad \left(\ii F^{--}(X,p)\right)^*=\ii F^{++}(X,p),\nonumber\\
\left(F^R(x,y)\right)^*&=&F^A(y,x)
\quad\hspace*{3.5mm}\Rightarrow\quad \left(F^R(X,p)\right)^*=F^A(X,p),
\end{eqnarray}
%
where the right parts specify the corresponding properties in the Wigner
representation. Diagrammatically these rules imply the simultaneous swapping
of all $+$ vertices into $-$ vertices and vice versa together with reversing
the line arrow-sense of all propagator lines in the diagram.

\end{fmffile}

\end{article}

\begin{thebibliography}{99}
\itemsep0mm
\bibitem{Schw}
J. Schwinger, {\em J. Math. Phys.} {\bf 2} (1961), 407.
\bibitem{Kad62}
L. P. Kadanoff and G. Baym, ``Quantum Statistical
Mechanics'', Benjamin, NY, 1962.
\bibitem{Keld64}
L. P. Keldysh, {\em ZhETF} {\bf 47} (1964), 1515; Engl. transl., {\em
Sov. Phys. JETP} {\bf 
20} (1965), 1018.
\bibitem{LP}
E. M. Lifshiz and L. P. Pitaevskii, ``Physical Kinetics'',
Pergamon press, 1981.
\bibitem{Land}
N. P. Landsman and Ch. G. van Weert, {\em Phys. Rep.} {\bf 145} (1987), 141.
\bibitem{BI00}
J. P. Blaizot and E. Iancu, hep-ph/0101103.
\bibitem{Dan84}
P. Danielewicz,{\em  Ann. Phys. (N.Y.)} {\bf 152} (1984), 239, and 305. 
\bibitem{Dan90}
P. Danielewicz,{\em  Ann. Phys. (N.Y.)} {\bf 197} (1990), 154.
\bibitem{Toh}
M. Tohyama, {\em Phys. Rev. C} {\bf 86} (1987), 187.
\bibitem{Bot90} 
W. Botermans and R. Malfliet,{\em  Phys. Rep.} {\bf 198} (1990), 115.
\bibitem{MSTV}  
A. B. Migdal, E. E. Saperstein, M. A. Troitsky and
D. N.Voskresensky,{\em  Phys. Rep.}  {\bf 192} (1990), 179.
\bibitem{Vos93}
D. N. Voskresensky,{\em  Nucl. Phys. A} {\bf 555} (1993), 293.
\bibitem{Hen}
P.A. Henning, {\em Phys. Rep. C} {\bf 253} (1995),
235.
\bibitem{Knoll95} 
J. Knoll and D. N. Voskresensky, {\em Phys. Lett. B}
{\bf 351} (1995), 43;\\
J. Knoll and D. N. Voskresensky, {\em Ann. Phys. (N.Y.)} {\bf 249} (1996),
532.
\bibitem{IKV}
Yu. B. Ivanov, J. Knoll, and D. N. Voskresensky,  
{\em Nucl. Phys. A} {\bf 657} (1999), 413.
\bibitem{Bozek98}
P. Bozek, {\em  Phys. Rev. C} {\bf 59} (1999),
2619. 
\bibitem{Knoll98}
J. Knoll,{\em  Prog. Part. Nucl. Phys.} {\bf 42} (1999), 177.  
\bibitem{IKV99}
Yu. B. Ivanov, J. Knoll, and D. N. Voskresensky,  
{\em Nucl. Phys. A} {\bf 672} (2000), 313.
\bibitem{Cass99}
W. Cassing and S. Juchem, {\em Nucl. Phys. A}
{\bf 665} (2000), 377; {\bf 672} (2000), 417.  
\bibitem{Leupold}
S. Leupold,{\em  Nucl. Phys. A} {\bf 672} (2000), 475. 
\bibitem{Mosel}
M. Effenberger, U. Mosel, {\em Phys. Rev. C} {\bf 60} (1999), 51901.
\bibitem{IKHV00}
Yu. B. Ivanov, J. Knoll, H. van Hees and D. N. Voskresensky,
e-Print Archive: nucl-th/0005075; {\em Yad. Fiz.} {\bf 64} (2001).
\bibitem{HK00}
H. van Hees, and J. Knoll, {\em Nucl. Phys. A} {\bf A683} (2001), 369. 
\bibitem{VS87}
D. N. Voskresensky and A. V. Senatorov,  {\em Yad. Fiz.} {\bf 45} (1987),
657; Engl. transl., {\em Sov. J. Nucl. Phys.} {\bf
45} (1987), 414.
\bibitem{Keil}
W. Keil, {\em Phys. Rev. D} {\bf 38} (1988), 152.
\bibitem{CalHu}
E. Calzetta and B. L. Hu, {\em Phys. Rev. D} {\bf 37} (1988),
2878.
\bibitem{Manson}
M. M\"{a}nson, and A. Sj\"{o}lander, {\em Phys. Rev. B} {\bf 11} (1975), 4639.
\bibitem{Korenman}
V. Korenman, {\em Ann. Phys. (N.Y.)} {\bf 39} (1966), 72.
\bibitem{Bez}
B. Bezzerides and D. F. DuBois,{\em  Ann. Phys. (N.Y.)} {\bf 70} (1972),
10.
\bibitem{Kraft}
W. D. Kraeft, D. Kremp, W. Ebeling and G. R\"{o}pke, ``Quantum
Statistics of Charged Particle Systems'', Akademie-Verlag, Berlin, 1986.
\bibitem{SerRai}
 J. W. Serene and D. Rainer, {\em Phys. Rep.} {\bf 101} (1983),
221. 
\bibitem{Chou} 
K. Chou, Z. Su, B. Hao and L. Yu, {\em Phys. Rep.}  {\bf 118} (1985), 1.
\bibitem{Rammer} 
J. Rammer and H. Smith, {\em Rev. Mod. Phys.} {\bf
58} (1986), 323.
\bibitem{Fauser}
R. Fauser, {\em Nucl. Phys. A} {\bf 606} (1996), 479.
\bibitem{LipS}  
V. Spicka and P. Lipavsky, {\em Phys. Rev. Lett.}
{\bf 73} (1994), 3439;  {\em Phys. Rev. B} {\bf 52} (1995), 14615.
\bibitem{Noziers}
P. Nozi\`ers, and E. Abrahams, {\em Phys. Rev. B} {\bf 10} (1974), 4932;\\
S. Abraham - Ibrahim, B. Caroli, C. Cardi, and B. Roulet, {\em Phys. Rev. B}
{\bf 18} (1978), 6702.
\bibitem{KadB} G. Baym and L. P. Kadanoff, {\em Phys. Rev.} {\bf 124} (1961),
287. 
\bibitem{Baym}
G. Baym, {\em Phys. Rev.} {\bf 127} (1962), 1391.
\bibitem{Luttinger}
J. M. Luttinger and J. C. Ward, {\em Phys. Rev.} {\bf 118} (1960), 1417.
\bibitem{Abrikos}
A. A. Abrikosov, L. P. Gorkov, I. E. Dzyaloshinski, 
``Methods of Quantum Field Theory in Statistical Physics'',
Dover Pub., INC. N.Y., 1975.
\bibitem{CJT} J. M. Cornwall, R. Jackiw and E. Tomboulis,
{\em Phys. Rev. D} {\bf 10} (1974), 2428.
\bibitem{LSV}
P. Lipavsky, V. Spicka and B. Velicky, {\em Phys. Rev. B} {\bf 34} (1986),
6933.
\bibitem{Bonitz} 
M. Bonitz, ``Quantum Kinetic Theory'', Teubner, Stuttgart/Leipzig, 1998.
\bibitem{Bornath}
Th. Bornath, D. Kremp, W. D. Kr\"aft, and M. Schlanges, {\em Phys. Rev. E}
{\bf 54} (1996), 3274. 
\bibitem{SCFNW}
M. Sch\"{o}nhofen, M. Cubero, B. Friman, W. N\"{o}renberg and Gy. Wolf,
{\em Nucl. Phys. A} {\bf 572} (1994), 112.
\bibitem{Jeon}
S. Jeon and L.G. Yaffe, {\em Phys. Rev. D} {\bf 53} (1996), 5799. 
\bibitem{VBRS}
D. N. Voskresensky, D. Blaschke, G. R\"{o}pke and H. Schulz,
{\em Int. Mod. Phys. J. E} {\bf 4} (1995), 1.
\bibitem{DB91} P. Danielewicz and G. Bertsch, {\em Nucl. Phys. A} {\bf 533}
 (1991), 712.
\bibitem{Riedel} 
E. Riedel, {\em Z. Phys. A} {\bf 210} (1968), 403.
\bibitem{Carneiro}
G. M. Carneiro and C. J. Pethick, {\em Phys. Rev. B} {\bf 11} (1975), 1106.
\bibitem{Baym91}
G. Baym and C. Pethick, ``Landau Fermi--Liquid Theory'', John
 Wiley and Sons, INC, N.Y., 1991.
\bibitem{Cass00}
W. Cassing and S. Juchem,{\em Nucl. Phys. A} {\bf 677} (2000), 445. 
\bibitem{Bozek}
P. Bo\'{z}ek and P. Czerski, nucl-th/0102020; 
P. Bo\'{z}ek, {\em Phys. Rev. C} {\bf 59} (1999), 2619;
{\em Nucl. Phys. A} {\bf 657} (1999), 187.
\bibitem{Dickhoff}
W. H. Dickhoff,{\em  Phys. Rev. C} {\bf 58} (1998), 2807; 
W. H. Dickhoff {\em et al.}, {\em Phys. Rev. C} {\bf 60} (1999), 4319. 
\bibitem{Dewulf}
Y. Dewulf, D. Van Neck and M. Waroquier, nucl-th/0012022. 
\bibitem{Lenz} W. Lenz, {\em Z. Phys.} {\bf 56} (1929), 778.  
\bibitem{Dover71}C. B. Dover, J. H\"ufner and R. H. Lemmer, {\em
Ann. Phys. (N.Y.)} {\bf 66} (1971), 248. 
\bibitem{DP}P. Danielewicz and S. Pratt, {\em Phys. Rev. C} {\bf 53} (1996),
249. 
\bibitem{BethU}
  E. Beth, G. E. Uhlenbeck, {\em Physica} {\bf 4} (1937), 915.  
\bibitem{Huang} K. Huang,
  "Statistical Mechanics", Wiley, New York (1963).  
\bibitem{DMB} R. Dashen,
  S. Ma, H. J. Bernstein, {\em Phys. Rev.} {\bf 187} (1969), 345.  
\bibitem{Mekjian} A. Z.   Mekjian, {\em Phys. Rev. C} {\bf 17} (1978), 1051.  
\bibitem{VPrakash} R.
  Venugopalan and M. Prakash, {\em Nucl. Phys. A} {\bf 456} (1992), 718.
\bibitem{Hees-Thesis}
H. van Hees, Ph-D thesis, Tu-Darmstadt, Germany, Landes und
Hochschulbibliothek, Oct. 2000, http://elib.tu-darmstadt.de/diss/000082/, to
be published in parts. 
\bibitem{HK01}H. van Hees and J. Knoll, to be published.
\end{thebibliography}
\end{document}